\documentclass[aps,pre,reprint,superscriptaddress,nopacs,nofootinbib]{revtex4-1}

\usepackage{amsmath,amssymb,latexsym}
\usepackage{graphicx}
\graphicspath{{../fig/}}
\usepackage{hyperref}
\usepackage{txfonts}
\usepackage{mathrsfs}
\usepackage{amscd}
\usepackage{color}
\DeclareSymbolFont{symbols}{OMS}{cmsy}{m}{n}
\DeclareSymbolFont{largesymbols}{OMX}{cmex}{m}{n}

\begin{document}

\title{
Fluctuation theorem for quantum-state statistics
}

\author{Naoto Tsuji}
\affiliation{RIKEN Center for Emergent Matter Science (CEMS), Wako 351-0198, Japan}
\author{Masahito Ueda}
\affiliation{Department of Physics, University of Tokyo, Hongo, Tokyo 113-0033, Japan}
\affiliation{RIKEN Center for Emergent Matter Science (CEMS), Wako 351-0198, Japan}

\begin{abstract}
We derive the fluctuation theorem for quantum-state statistics that can be obtained when
we initially measure the total energy of a quantum system at thermal equilibrium,
let the system evolve unitarily,
and record the quantum-state data reconstructed at the end of the process.
The obtained theorem shows that 
the quantum-state statistics for the forward and backward processes 
is related to the equilibrium free-energy difference
through an infinite series of independent relations,
which gives the quantum work fluctuation theorem as a special case,
and reproduces the out-of-time-order fluctuation-dissipation theorem near thermal equilibrium.
The quantum-state statistics exhibits a system-size scaling behavior that differs
between integrable and non-integrable (quantum chaotic) systems
as demonstrated numerically for one-dimensional quantum lattice models.
\end{abstract}


\date{\today}

\maketitle

Fluctuation theorems (FTs) have played a central role
in our understanding of how macroscopic irreversibility arises from
microscopically reversible equation of motion
\cite{BochkovKuzovlev1977,EvansCohenMorriss1993,GallavottiCohen1995,Jarzynski1997,Crooks1999,
EspositoHarbolaMukamel2009,Campisi2011}.
The FTs lead to many fundamental relations
in thermodynamics and statistical mechanics,
including the second law of thermodynamics,
the fluctuation-dissipation theorem (FDT)
\cite{CallenWelton1951,Kubo1957,Marconi2008}, 
and Onsager's reciprocity relation
 \cite{Onsager1931,Casimir1945}.


The conventional approach to FTs in isolated quantum systems
is based on the two-point measurement for work \cite{Kurchan2000,Tasaki2000,EspositoHarbolaMukamel2009}:
one initially measures the total energy,
let the system evolve according to a time-dependent Hamiltonian, 
and again measures the total energy at the end of the process.
From the difference between the initial and final total energies, 
one can extract the work done on the system by an external force.
The obtained work probability distributions for the forward and time-reversed processes
are related to the equilibrium free-energy difference between the initial and final configurations
(the quantum work FT).
In this approach, one makes a projective energy measurement (with the outcome $E_l^f$ being the $l$th eigenenergy
of the final Hamiltonian)
on the final state $\hat\rho$,
so that one obtains limited information on the quantum state $\hat\rho$ itself,
i.e., only the diagonal information $\langle E_l^f|\hat\rho|E_l^f\rangle$
is available, where $|E_l^f\rangle$ is the energy eigenstate.

\begin{figure}[t]
\includegraphics[width=8.5cm]{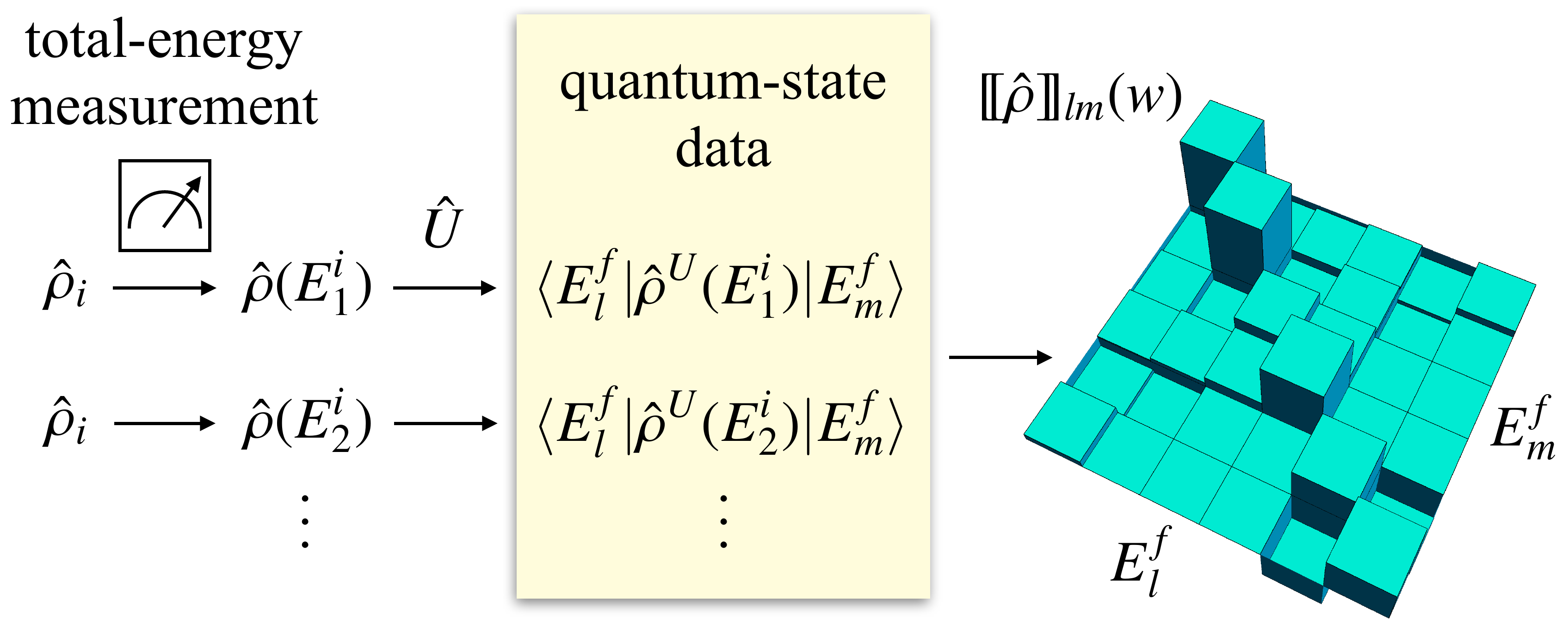}
\caption{Schematic procedure for measuring quantum-state statistics. 
We initially prepare the thermal equilibrium state $\hat\rho_i$,
and measure the total energy with the outcome $E_k^i$, where the state changes to $\hat\rho(E_k^i)$.
Then the system evolves according to a unitary operator $\hat U$, and the final state $\hat\rho^U(E_k^i)$
is reconstructed in the energy eigenbasis.
We repeat the procedure to accumulate the matrix data $[\![\hat\rho]\!]_{lm}(w)$ [Eq.~(\ref{qss})].}
\label{quantum-state statistics}
\end{figure}

How does the quantum state $\hat\rho$ realized after the time evolution (including information
on the off-diagonal elements $\langle E_l^f|\hat\rho|E_m^f\rangle$, $l\neq m$) fluctuate?
Here, by fluctuations of the quantum state we mean that the state fluctuates
depending on the result of the initial energy measurement.
If we repeat the procedure to (i) prepare the initial thermal equilibrium state,
(ii) measure the total energy, (iii) perform a unitary time evolution, 
and (iv) reconstruct the quantum state $\hat\rho$, 
we can operationally determine {\it the statistics of quantum states} (Fig.~\ref{quantum-state statistics}).
When the above procedure is repeated sufficiently many times,
we obtain duplicated copies of quantum states,
with which we can in principle reconstruct the quantum state using the technique of the quantum-state tomography
\cite{ParisRehacekBook,LvovskyRaymer2009}.

The statistics of quantum states
is closely related to
quantum chaos, or non-integrability, of the system,
the characterization of which has been a long-standing issue in statistical mechanics
\cite{Berry1987,HaakeBook}.
Suppose that after the first measurement
the quantum state is projected to a certain eigenstate of the initial Hamiltonian.
Then the state evolves within a subspace of the total Hilbert space
due to the presence of conserved quantities. For integrable systems,
the number of conserved quantities is extensive, so that the size of the subspace
is highly constrained. Hence we expect that the resulting behavior of the quantum-state statistics is different
between integrable and non-integrable systems. 

Another motivation to study the quantum-state statistics is the recent finding of
the out-of-time-order FDT \cite{TsujiShitaraUeda2018},
which relates chaotic properties of the system and a nonlinear response function involving a time-reversed process,
and can be viewed as a higher-order extension of the conventional FDT.
Provided that the conventional FDT can be derived from the quantum work FT near equilibrium,
it is thus a natural question what is the underlying law that leads to the out-of-time-order
FDT if applied near equilibrium.

In this paper, we show that the quantum-state statistics accumulated under a certain condition
for the forward and time-reversed processes
satisfies an infinite series of exact relations that are expressed in terms of the equilibrium free-energy difference 
between the initial and final configurations.
The relations include the quantum work (Crooks) FT as a special case, and allow further extensions.
Near equilibrium, the out-of-time-order FDT \cite{TsujiShitaraUeda2018} is reproduced.
We argue that the fluctuation of the quantum-state statistics
shows a different system-size scaling between integrable and non-integrable systems,
which can be used as a diagnosis of quantum chaos.
This is demonstrated numerically for one-dimensional quantum lattice models.


Let us suppose that an isolated quantum system evolves in time according to the time-dependent Hamiltonian $\hat H(s)$ 
($t_i\le s\le t_f$)
(forward process).
The initial and final Hamiltonians are denoted by $\hat H_i=\hat H(t_i)$ and $\hat H_f=\hat H(t_f)$.
The unitary evolution operator is given by $\hat U=\mathcal T \exp(-\frac{i}{\hbar}\int_{t_i}^{t_f} ds\,\hat H(s))$,
where $\mathcal T$ represents the time-ordered product.
We assume that the initial state is in thermal equilibrium with temperature $k_BT=\beta^{-1}$, 
and is described by the canonical ensemble with 
the density matrix $\hat\rho_i=e^{-\beta\hat H_i}/Z_i(\beta)$,
where $Z_i(\beta)\equiv {\rm Tr}(e^{-\beta\hat H_i})$ is the partition function.
We denote the eigenvalues and orthonormal eigenvectors of $\hat H_i$ ($\hat H_f$)
by $E_k^i$ ($E_k^f$) and $|E_k^i\rangle$ ($|E_k^f\rangle$), respectively.

Suppose that we perform a projective energy measurement and obtain the measurement outcome $E_k^i$ with the probability $p_k^i=e^{-\beta E_k^i}/Z_i(\beta)$,
where the quantum state $\hat\rho_i$ is projected from $\hat\rho_i$ to $\hat\rho(E_k^i)=|E_k^i\rangle \langle E_k^i|$.
After the unitary time evolution,
the quantum state becomes $\hat\rho^U(E_k^i)\equiv\hat U \hat\rho(E_k^i) \hat U^\dagger$.
At the end of the process,
we record the quantum state reconstructed in the eigenbasis of the final Hamiltonian as
$\langle E_l^f| \hat\rho^U(E_k^i) |E_m^f\rangle$.
We here address the question of whether
there is any law that governs the statistics of these quantum-state data
when we repeat the above procedure.
We show that it emerges when we accumulate the quantum-state data
under a certain energy constraint given by $w=\frac{1}{2}(E_l^f+E_m^f)-E_k^i$.
After taking the average, we obtain
\begin{align}
&
[\![ \hat\rho]\!]_{lm}(w)
\equiv
\overline{\delta(w-(\tfrac{1}{2}(E_l^f+E_m^f)-E_k^i))
\langle E_l^f| \hat\rho^U(E_k^i) |E_m^f\rangle}
\notag
\\
&=
\sum_k p_k^i \delta(w-(\tfrac{1}{2}(E_l^f+E_m^f)-E_k^i))
\langle E_l^f| \hat\rho^U(E_k^i) |E_m^f\rangle,
\label{qss}
\end{align}
where the overline represents the average over the repeated processes, and $\delta(x)$ is the Dirac delta function.
For $l=m$, $w$ corresponds precisely to the difference between the initial 
and final energies, which is equivalent to the work performed on the system. 
However, for off-diagonal elements, $w$ does not, in general, correspond to the work,
but only has a formal meaning of the difference between the initial energy $E_k^i$
and the averaged final energy $\frac{1}{2}(E_l^f+E_m^f)$.

We also consider the time-reversed process
with the Hamiltonian $\hat{\bar H}(s)=\Theta\hat H(t_i+t_f-s)\Theta^{-1}$ ($t_i\le s\le t_f$), where $\Theta$ represents the antiunitary time-reversal operator.
The corresponding initial and final Hamiltonians are $\hat{\bar H}_i=\hat{\bar H}(t_i)$ and $\hat{\bar H}_f=\hat{\bar H}(t_f)$,
and the unitary evolution is given by $\hat{\bar U}=\Theta\hat U^\dagger\Theta^{-1}$. The initial state for the time-reversed process
is assumed to be $\hat{\bar\rho}_i=e^{-\beta\hat{\bar H}_i}/\bar Z_i(\beta)$, 
where $\bar Z_i(\beta)={\rm Tr}(e^{-\beta\hat{\bar H}_i})$. In the same way as the forward process, we define
\begin{align}
&
[\![ \hat{\bar\rho}]\!]_{lm}(w)
\equiv
\overline{\delta(w-(\tfrac{1}{2}(\bar E_l^f+\bar E_m^f)-\bar E_k^i))
\langle \bar E_l^f| \hat{\bar\rho}^{\bar U}(\bar E_k^i) |\bar E_m^f\rangle}
\notag
\\
&=
\sum_k \bar p_k^i \delta(w-(\tfrac{1}{2}(\bar E_l^f+\bar E_m^f)-\bar E_k^i))
\langle \bar E_l^f| \hat{\bar\rho}^{\bar U}(\bar E_k^i) |\bar E_m^f\rangle,
\end{align}
where $\bar E_k^i$ ($\bar E_k^f$) and $|\bar E_k^i\rangle$ ($|\bar E_k^f\rangle$) are the eigenvalues and orthonormal eigenvectors 
of $\hat{\bar H}_i$ ($\hat{\bar H}_f$),
respectively, $\bar p_k^i=e^{-\beta \bar E_k^i}/\bar Z_i(\beta)$,
$\hat{\bar\rho}^{\bar U}(\bar E_k^i)=\hat{\bar U}\hat{\bar\rho}(\bar E_k^i)\hat{\bar U}^\dagger$,
and $\hat{\bar\rho}(\bar E_k^i)=|\bar E_k^i\rangle \langle \bar E_k^i|$.

Since $[\![\hat\rho]\!](w)$ is an operator acting on the Hilbert space,
there are various ways to retrieve information from this object. 
Let us define distribution functions for the quantum-state statistics by taking the trace of the $n$th moment of $[\![\hat\rho]\!](w)$
($n=1,2,\dots$),
\begin{align}
p_n(w)
&\equiv
\frac{1}{\mathcal N_n}{\rm Tr}([\![\hat\rho]\!]^{\circledast n}(w)).
\label{p_n}
\end{align}
Here $\mathcal N_n$ is a normalization constant determined by
\begin{align}
\int_{-\infty}^{\infty} dw\, p_n(w)
&=1,
\label{normalization}
\end{align}
and $[\![\hat\rho]\!]^{\circledast n}(w)=([\![\hat\rho]\!]\circledast \cdots \circledast [\![\hat\rho]\!])(w)$ is defined by 
the $n$th power of $[\![\hat\rho]\!](w)$
with the symbol $\circledast$ denoting the matrix multiplication and energy convolution simultaneously, i.e.,
\begin{align}
([\![\hat\rho]\!]\circledast [\![\hat\rho]\!])_{lm}(w)
&\equiv
\int_{-\infty}^{\infty} dw' \sum_n [\![\hat\rho]\!]_{ln}(w-w')
[\![\hat\rho]\!]_{nm}(w').
\end{align}
For the time-reversed process, the corresponding distribution function is defined by
$\bar p_n(w)\equiv\frac{1}{\bar{\mathcal N}_n}{\rm Tr}([\![\hat{\bar\rho}]\!]^{\circledast n}(w))$
with the normalization condition $\int_{-\infty}^{\infty} dw\, \bar p_n(w)=1$
and $\bar{\mathcal N}_n$ being the normalization constant for $\bar p_n(w)$.

At $n=1$, $p_n(w)$ is identical to the work probability distribution: 
$p_1(w)=\sum_{kl} p_k^i \delta(w-E_l^f+E_k^i) |\langle E_l^f|\hat U|E_k^i\rangle|^2$.
For arbitrary $n$, $p_n(w)$ can be proven to take a real value (Appendix~\ref{appendix A}).
However, for $n\ge 2$, $p_n(w)$ is not necessarily positive semidefinite.
This prevents us from interpreting $p_n(w)$ ($n\ge 2$) as a probability distribution, though
$p_n(w)$ satisfies the normalization condition (\ref{normalization}).
Hence $p_n(w)$ ($n\ge 2$) should be regarded as a quasiprobability.

The main result of this paper is that the following relation holds between $p_n(w)$ and 
its time-reversed partner $\bar p_n(w)$:
\begin{align}
\frac{p_n(w)}{\bar p_n(-w)}
&=
e^{\beta(w-n\Delta F(n\beta))}
\quad
(n=1,2,\dots).
\label{qs ft}
\end{align}
Here $\Delta F(\beta)=F_f(\beta)-F_i(\beta)$ [$F_{i,f}(\beta)=-\beta^{-1}\ln Z_{i,f}(\beta)$] is the difference
of the equilibrium free energies for the initial and final Hamiltonians at temperature $\beta^{-1}$.
Note that the inverse temperature appearing in the free-energy argument is multiplied by $n$ in Eq.~(\ref{qs ft}).
For $n=1$, the relation (\ref{qs ft}) reduces to the quantum work FT,
$p_1(w)/\bar p_1(-w)=e^{\beta(w-\Delta F(\beta))}$. For $n\ge 2$, the relation (\ref{qs ft}) gives 
an extension of the FT to the quantum-state statistics.
A remarkable feature of Eq.~(\ref{qs ft}) is that it is valid for arbitrary unitary evolution $\hat U$,
no matter how the system is driven away from equilibrium.
Note that on the left-hand side of Eq.~(\ref{qs ft}) each $p_n(w)$ and $\bar p_n(-w)$
strongly depends on $\hat U$, while the right-hand side is written in terms of the equilibrium quantities.

The relation (\ref{qs ft}) can be derived using the method of characteristic
functions \cite{Talkner2007}. 
Here we define a characteristic function for $p_n(w)$ as the Fourier transform of $p_n(w)$,
\begin{align}
G_n(u)
&\equiv
\int_{-\infty}^{\infty} dw\, e^{iuw} p_n(w),
\label{G_n}
\end{align}
which can be written as
$G_n(u)=\mathcal N_n^{-1}{\rm Tr}[(\hat\rho_i\hat W_{i,u}^\dagger(t_i)\hat W_{f,u}(t_f)^n)]$,
where $\hat W_{i,u}(t_i)$ and $\hat W_{f,u}(t_f)$ are the Heisenberg representation
of operators $\hat W_{i,u}\equiv e^{iu\hat H_i}$ and $\hat W_{f,u}\equiv e^{iu\hat H_f}$, respectively
(Appendix~\ref{appendix A}).
Hence $G_n(u)$ ($n\ge 2$) is classified into an out-of-time-ordered correlation function \cite{LarkinOvchinnikov1969}.
By using the time-reversal property of $G_n(u)$, we find a symmetry relation 
$G_n(u)=(Z_f(n\beta)/Z_i(n\beta))\bar G_n(-u+i\beta)$, 
where $\bar G_n(u)$ is the characteristic function for $\bar p_n(w)$.
After Fourier transformation, we arrive at Eq.~(\ref{qs ft}). The details of the proof is described in Appendix~\ref{appendix A}.

By multiplying $e^{-\beta w}\bar p_n(-w)$ on both sides of Eq.~(\ref{qs ft})
and using the normalization condition (\ref{normalization}), we obtain
the integral FT for the quantum-state statistics,
\begin{align}
\langle e^{-\beta w}\rangle_{p_n}
&=
e^{-n\beta\Delta F(n\beta)},
\label{qs Jarzynski}
\end{align}
where $\langle \cdots \rangle_{p_n}\equiv \int_{-\infty}^{\infty} dw\, p_n(w)\cdots$.
For $n=1$, the relation (\ref{qs Jarzynski}) is nothing but the Jarzynski equality, 
$\langle e^{-\beta w}\rangle_{p_1}=e^{-\beta\Delta F(\beta)}$, while for $n\ge 2$
it provides an extension of the Jarzynski equality.
If one knows the distribution function $p_n(w)$, one can extract the equilibrium free-energy difference
at temperature $k_BT/n=(n\beta)^{-1}$.
Since $p_n(w)$ is generated by the characteristic function $G_n(u)$,
one can measure $p_n(w)$ through the measurement of the out-of-time-ordered correlation function,
for which various protocols have been proposed
\cite{Swingle2016,Campisi2016,Yao2016,Zhu2016,TsujiWernerUeda2016,TsujiShitaraUeda2018,
YungerHalpern2017,YungerHalpernSwingleDressel2017}.

Applying Jensen's inequality 
to the Jarzynski equality,
one arrives at the second law of thermodynamics, 
\begin{align}
\langle w \rangle_{p_1}
&\ge
\Delta F(\beta).
\label{second law}
\end{align}
One may wonder if one could derive a similar inequality
\begin{align}
\langle w\rangle_{p_n}
\ge
n\Delta F(n\beta)
\quad
(!)
\label{qs second law}
\end{align}
from Eq.~(\ref{qs Jarzynski}). This is, however, possible only if $p_n(w)$ is positive semidefinite,
since one cannot use Jensen's inequality for non-positive-semidefinite distributions.
We note that $p_n(w)$ becomes positive semidefinite in
the zero-temperature limit ($\beta\to\infty$). Let us assume that
the ground state of the initial system (denoted by $|E_g^i\rangle$ with the eigenenergy $E_g^i$) is unique. 
Then, in the zero-temperature limit,
\begin{align}
p_n(w)
&\to
\frac{1}{\mathcal N_n}\sum_{l_1,\dots, l_n} \delta(w-(E_{l_1}^f+\cdots+E_{l_n}^f)+nE_g^i)(p_g^i)^n
\notag
\\
&\quad\times
|\langle E_{l_1}^f|\hat U|E_g^i\rangle|^2
\cdots
|\langle E_{l_n}^f|\hat U|E_g^i\rangle|^2
\ge 0
\end{align}
with $p_g^i=e^{-\beta E_g^i}/Z_i(\beta)$.
Thus, at zero temperature the inequality (\ref{qs second law}) holds.
Of course, this does not mean that we have a new second law in addition to the existing one (\ref{second law}).
At zero temperature $p_n(w)$ is related to $p_1(w)$ through
$p_n(w)=\int_{-\infty}^{\infty} dw_1 \cdots dw_{n-1}\, p_1(w-w_1)p_1(w_1-w_2)\cdots p_1(w_{n-2}-w_{n-1})p_1(w_{n-1})$, 
from which one obtains
$\langle w\rangle_{p_n}=n\langle w\rangle_{p_1}$. Therefore, the inequality (\ref{qs second law})
reduces to the second law (\ref{second law}) at zero temperature [where $\Delta F(n\beta)\sim \Delta F(\beta)$], 
and (\ref{qs second law}) does not provide new information in this case. 
In fact, the relation (\ref{qs ft}) reduces to the quantum work FT [Eq.~(\ref{qs ft}) with $n=1$] 
in the zero-temperature limit.
To obtain new information beyond the quantum work FT,
one has to consider finite-temperature states.

If the relation (\ref{qs ft}) is applied near equilibrium,
one can reproduce the out-of-time-order FDT \cite{TsujiShitaraUeda2018}
around zero frequency.
This can be seen from the expansion of the integral FT (\ref{qs Jarzynski}) for $n=1$ and $n=2$
up to the third cumulants with respect to $w$.
If the Hamiltonian is split into the time-independent part and the rest as
$\hat H(s)=\hat H_0+\xi(s)\hat X(s)$, 
where $\xi(s)$ is an external field and $\hat X(s)$ is the coupled operator,
then the second-order functional derivative $\frac{\delta^2}{\delta\xi(s)\xi(s')}$
on both sides of the cumulant expansions around $\xi(s)=0$ (near equilibrium) leads to
the near-zero-frequency part of the out-of-time-order FDT.
Details of the derivation are given in Appendix~\ref{appendix B}.

We have examined two aspects of $p_n(w)$: the distribution function for the quantum-state statistics
and out-of-time-ordered correlation functions. For the latter, there have been various discussions
in relation to chaotic properties of quantum systems
\cite{Kitaev2015,MaldacenaShenkerStanford2016,Swingle2016,Rozenbaum2016,Aleiner2016,Fan2017,TsujiShitaraUeda2018b}.
Here we argue that there is a strong connection between the fluctuation in $p_n(w)$ ($n\ge 2$) and quantum chaotic nature
(non-integrability) of the system.
The crucial difference of $p_n(w)$ ($n\ge 2$) from the work probability distribution $p_1(w)$
is that the former can take a negative value.
In the following, we focus on the case of $n=2$.
We quantify the fluctuation in $p_2(w)$ by the $L^1$ norm ($\| \cdot \|_1$),
\begin{align}
\Delta p_2
&\equiv
\frac{1}{Z_i(\beta)}\| p_2(w) \|_1
=
\frac{1}{Z_i(\beta)}\int_{-\infty}^{\infty} dw\, |p_2(w)|.
\label{Delta p_2}
\end{align}
$\Delta p_2$ counts the negative portion of $p_2(w)$ since $\Delta p_2=Z_i(\beta)^{-1}[1-2\int_{p_2(w)<0} dw\, p_2(w)]$
(note that $p_2(w)$ satisfies the normalization condition (\ref{normalization})).

As an illustration, let us consider the case that the Hamiltonian is suddenly quenched (i.e., $\hat H(s)=\hat H_i\to\hat H_f$) and the initial temperature is $\beta=0$. If we assume a non-degeneracy condition
(Appendix~\ref{appendix C}), $\Delta p_2$ is written for real Hamiltonians as
$\Delta p_2=
Z_i(0)^{-2} \sum_{klmn} 
|\langle E_k^i|E_n^f\rangle|
\cdot
|\langle E_n^f|E_m^i\rangle|
\cdot
|\langle E_m^i|E_l^f\rangle|
\cdot
|\langle E_l^f|E_k^i\rangle|
$.
Using conserved quantities inherent in the system,
the unitary transition matrix $\hat U_{lk}\equiv \langle E_l^f|E_k^i\rangle$
can be block-diagonalized as $\hat U=\oplus_\alpha \hat U^{(\alpha)}$.
If we define an entrywise-absolute-value matrix, $(\hat U_{\rm abs}^{(\alpha)})_{lk}\equiv |\hat U_{lk}^{(\alpha)}|$,
then $\Delta p_2=Z_i(0)^{-2}\sum_\alpha {\rm Tr}(\hat U_{\rm abs}^{(\alpha)\dagger}\hat U_{\rm abs}^{(\alpha)}\hat U_{\rm abs}^{(\alpha)\dagger} \hat U_{\rm abs}^{(\alpha)})
=Z_i(0)^{-2}\sum_\alpha \|\hat U_{\rm abs}^{(\alpha)\dagger}\hat U_{\rm abs}^{(\alpha)}\|_F^2$,
where $\|\cdot\|_F$ denotes the Frobenius norm. Since the Frobenius norm is submultiplicative,
$\Delta p_2$ satisfies an inequality,
$\Delta p_2\le Z_i(0)^{-2}\sum_\alpha \| \hat U_{\rm abs}^{(\alpha)\dagger}\|_F^2 \| \hat U_{\rm abs}^{(\alpha)}\|_F^2$.
By using the relation
$\|\hat U_{\rm abs}^{(\alpha)}\|_F^2=\|\hat U_{\rm abs}^{(\alpha)\dagger}\|_F^2=\|\hat U^{(\alpha)}\|_F^2=
{\rm Tr}(\hat U^{(\alpha)}\hat U^{(\alpha)\dagger})=D_\alpha$
($D_\alpha$ is the dimension of the $\alpha$th block Hilbert space) and $Z_i(0)=\sum_\alpha D_\alpha=D$
($D$ is the dimension of the total Hilbert space), we obtain
\begin{align}
\Delta p_2
&\le
\frac{\sum_\alpha D_\alpha^2}{(\sum_\alpha D_\alpha)^2}.
\label{Delta p_2 inequality}
\end{align}
The right-hand side of this inequality strongly depends on the number of conserved quantities.
As an estimate, let's suppose that each block Hilbert space has approximately the same dimension
(i.e., $D_\alpha$ is independent of $\alpha$).
Then $\Delta p_2 \lesssim D_\alpha/D$, i.e., the fluctuation in $p_2(w)$ is constrained by
the dimension of the block Hilbert space as compared to the dimension of the total Hilbert space.
In integrable systems, the number of conserved quantities typically grows in proportion to the system size,
so that $D_\alpha/D$ is expected to decay exponentially in the large system-size limit.
On the other hand, in non-integrable systems there is a finite number of conserved quantities,
so that $D_\alpha/D$ remains constant (or decays at most algebraically) as the system size increases.
One can thus distinguish integrable and non-integrable systems 
by examining the system-size scaling behavior of $\Delta p_2$.

\begin{figure}[t]
\includegraphics[width=8cm]{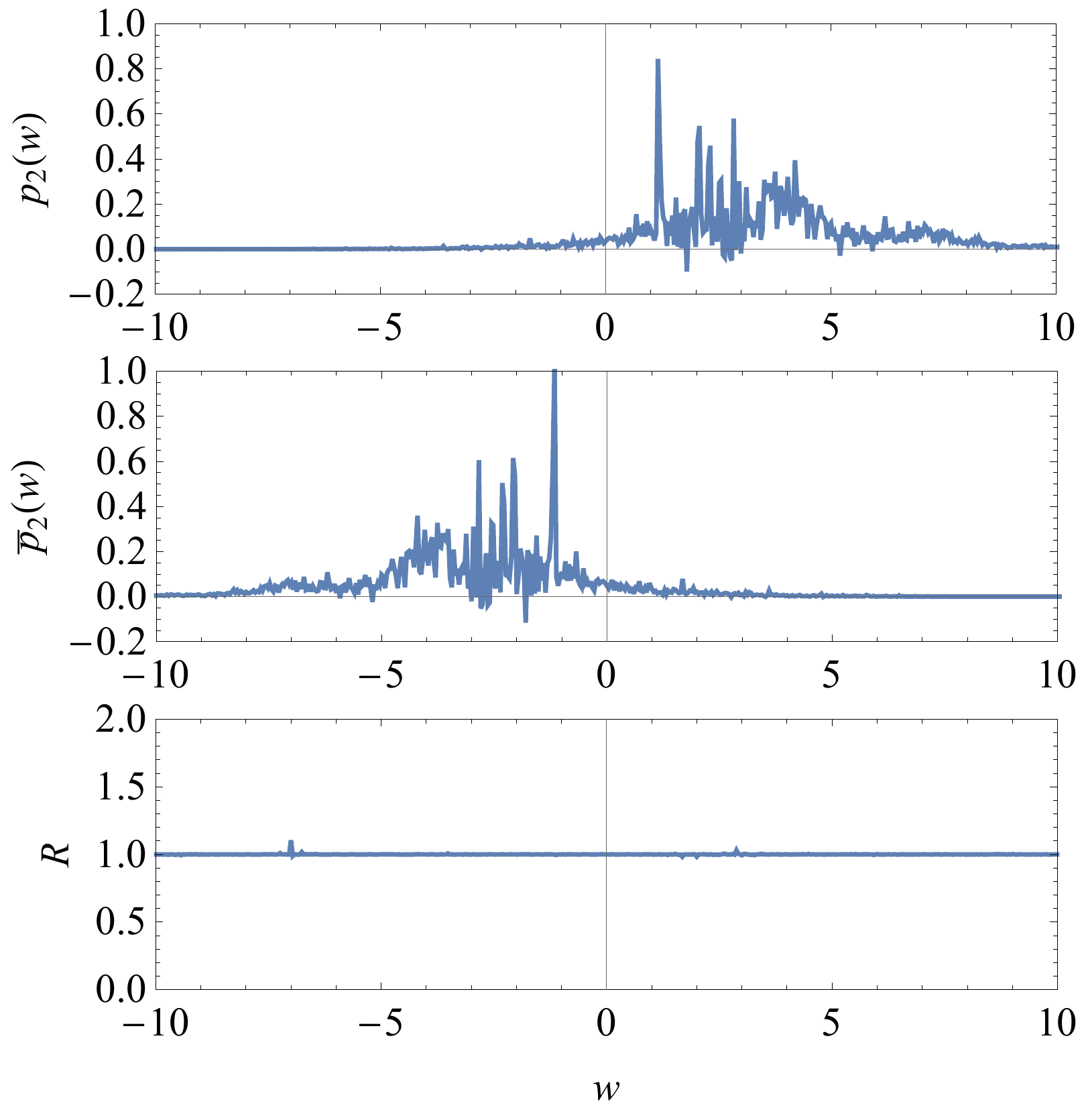}
\caption{Plot of $p_2(w)$ for the forward process (\ref{p_n}) (top panel)
and that of $\bar p_2(w)$ for the time-reversed process (middle) in the one-dimensional hard-core boson model (\ref{hardcore boson})
driven by the interaction quench $V=2\to 4$ with $t'=V'=1$, $\beta=0.1$, $L=12$, and $N=4$. 
The bottom panel plots $R=p_2(w)/\bar p_2(-w)
/e^{\beta(w-2\Delta F(2\beta))}$ as a function of $w$. The finite-size grid ($\Delta w=0.04$) is used.}
\label{fig:p_2}
\end{figure}

We numerically demonstrate the relation (\ref{qs ft}) for the quantum-state statistics and the behavior of $\Delta p_2$ (\ref{Delta p_2})
for the one-dimensional model of hard-core bosons with the Hamiltonian,
\begin{align}
\hat H(s)
&=
-t\sum_{i} (b_i^\dagger b_{i+1}+\mbox{h.c.})+V(s)\sum_i n_i^b n_{i+1}^b
\notag
\\
&\quad
-t'\sum_{i} (b_i^\dagger b_{i+2}+\mbox{h.c.})+V'\sum_i n_i^b n_{i+2}^b,
\label{hardcore boson}
\end{align}
where $t$ ($t'$) and $V(s)$ ($V'$) are the (next-)nearest-neighbor hopping and the strength of the interaction, respectively,
and $b_i^\dagger$ is the creation operator for hard-core bosons at site $i$.
We use $t$ as the unit of energy,
and assumes the periodic boundary condition.
The results are shown for the filling $N/L=1/3$,
where $N$ and $L$ are the number of particles and lattice sites, respectively.
For other fillings, we obtain qualitatively similar results (Appendix~\ref{appendix C}).
To drive the system out of equilibrium, we perform an interaction quench $V(s)=V_i \to V_f$ at time $s=0$.
In this setup, $p_n(w)$ (\ref{p_n}) does not depend on $t_i(<0)$ and $t_f(>0)$.
We numerically solve the model by exact diagonalization (for details, see Appendix~\ref{appendix C}).

The model (\ref{hardcore boson}) has been well studied in the context of quantum chaos \cite{Rigol2009,SantosRigol2010}.
At $t'=V'=0$, the model is known to be integrable.
In the non-integrable case ($t'\neq 0$ or $V'\neq 0$), the level-spacing statistics shows the Wigner-Dyson distribution,
which is the universal property of quantum chaotic systems as expected from random matrix theory.
The non-integrable model satisfies the eigenstate thermalization hypothesis
\cite{Deutsch1991,Srednicki1994,RigolDunjkoOlshanii2008},
which is a sufficient condition for an isolated quantum system to be thermalized.

In the top and middle panels in Fig.~\ref{fig:p_2}, we plot the distribution functions 
$p_2(w)$ for the forward process and $\bar p_2(w)$ for the time-reversed processes with $\beta=0.1$, 
where we take a finite grid size $\Delta w=0.04$ to broaden the delta function
(Appendix~\ref{appendix C}).
We clearly see that both $p_2(w)$ and $\bar p_2(w)$ have
negative parts.
In the bottom panel of Fig.~\ref{fig:p_2}, 
we plot $R\equiv p_2(w)/\bar p_2(-w)/e^{\beta(w-2\Delta F(2\beta))}$.
The value of $R$ stays close to $1$ over the whole region of $w$, which confirms the validity of the FT (\ref{qs ft})
for the quantum-state statistics.
Small derivations are due to the finite grid $\Delta w$ used to plot $p_2(w)$ and $\bar p_2(w)$.

\begin{figure}[t]
\includegraphics[width=8.5cm]{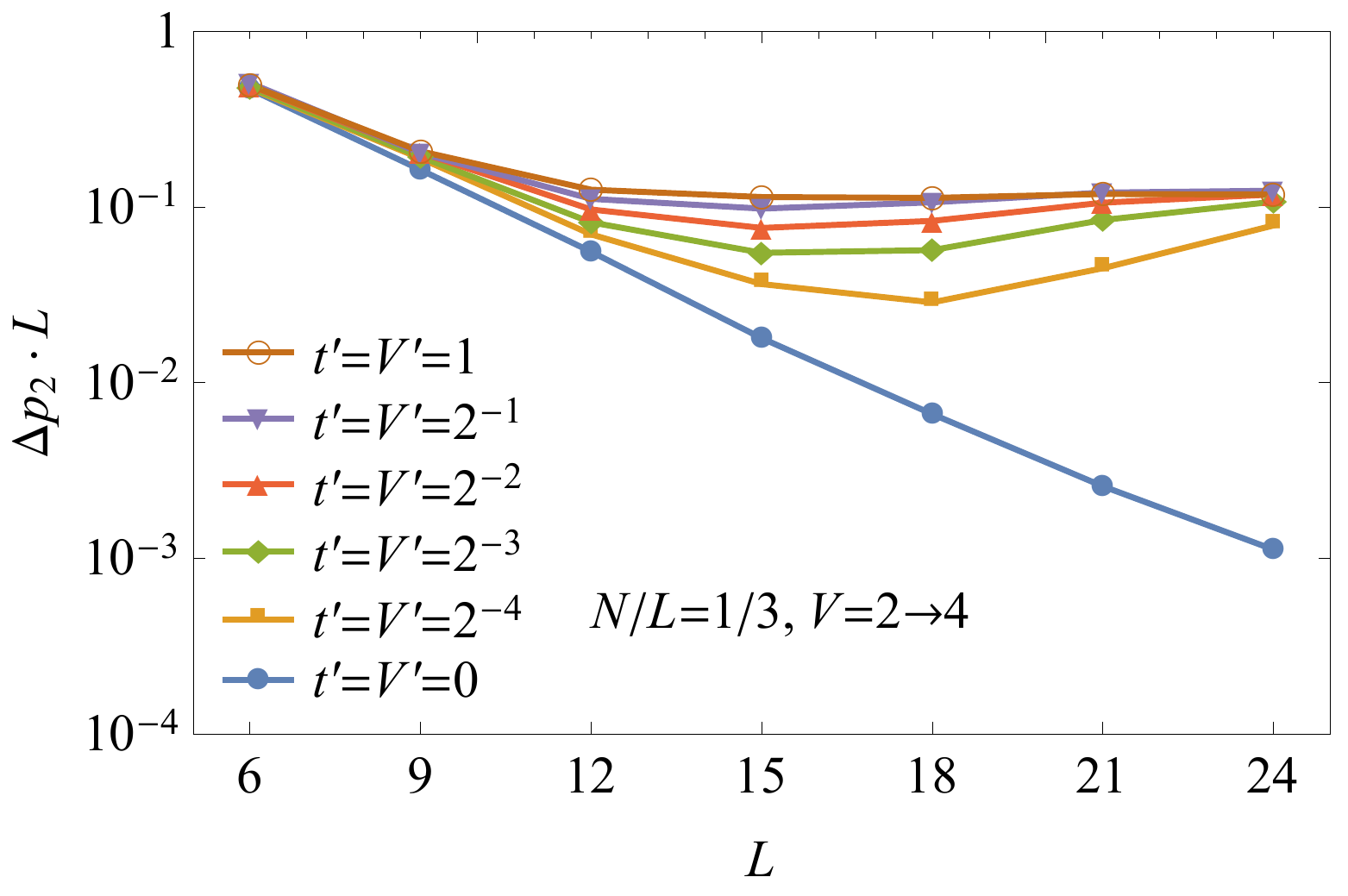}
\caption{Log plot of $\Delta p_2\cdot L$ against the system size $L$ 
for the one-dimensional hard-core boson model (\ref{hardcore boson}) with $\beta=0$
driven by the interaction quench $V=2\to 4$.
The system is integrable if $t'=V'=0$ and non-integrable otherwise.}
\label{fig:Delta p_2}
\end{figure}


We numerically evaluate $\Delta p_2$ (\ref{Delta p_2}), which quantifies the negative portion of the distribution $p_2(w)$,
for the one-dimensional hardcore boson model (\ref{hardcore boson})
in the limit of $\Delta w\to 0$ while keeping $L$ fixed (Appendix~\ref{appendix C}).
At zero temperature, $p_2(w)$ is positive semidefinite (i.e., $\Delta p_2=Z_i(\beta)^{-1}$) as explained earlier, 
and $\Delta p_2$ grows monotonically as temperature increases.
In Fig.~\ref{fig:Delta p_2}, we plot $\Delta p_2$ multiplied by the system size $L$ as a function of $L$ at $\beta=0$
for the quench $V=2\to 4$.
Clearly, $\Delta p_2$ shows a different scaling behavior
between the integrable ($t'=V'=0$) and non-integrable ($t'=V'\neq 0$) cases.
For the integrable case, $\Delta p_2$ tends to decay exponentially (within $L\le 24$ one can still see slight bending of the curve
in the log plot in Fig.~\ref{fig:Delta p_2}),
while for the non-integrable cases $\Delta p_2$ decays algebraically ($\Delta p_2\propto L^{-1}$)
and converges to the single universal curve.
Even a tiny violation of integrability ($t'=V'=2^{-4}$) causes a big difference in the behavior of $\Delta p_2$. 
These results are consistent with the inequality (\ref{Delta p_2 inequality}).
For the one-dimensional hardcore boson model (\ref{hardcore boson}), in the non-integrable case
$D=\binom{L}{N}$ and $D_\alpha\approx \frac{1}{L}\binom{L}{N}$
due to the parity and translational symmetries. From (\ref{Delta p_2 inequality}),
$\Delta p_2$ is roughly bounded by $\Delta p_2\lesssim L^{-1}$.
If $\Delta p_2$ decays as a power law, $\Delta p_2\propto L^{-\gamma}$, then the exponent $\gamma$ must satisfy
$\gamma\ge 1$.
The results shown in Fig.~\ref{fig:Delta p_2} indicate that the inequality for the exponent $\gamma$ is saturated
(i.e., $\gamma=1$).
In the integrable case shown in Fig.~\ref{fig:Delta p_2}, 
the numerical estimate within $L\le 24$ suggests that $\Delta p_2\propto e^{-c L}$ with $c=0.30$,
the value of which is, however, non-universal and depends on the model parameters.
We also simulate the same quantity for the one-dimensional spinless fermion model with nearest and next nearest neighbor
hopping and interaction \cite{Rigol2009b,SantosRigol2010}, and obtain similar results (Appendix~\ref{appendix C}).


To summarize, we have studied the statistics of quantum states
that can be obtained by the projective energy measurement followed by
unitary evolution and quantum-state reconstruction in the energy basis. 
By accumulating the data of quantum states under a certain energy condition [Eq.~(\ref{qss})],
we obtain the distribution function [Eq.~(\ref{p_n})] which satisfies an infinite series of exact relations [Eq.~(\ref{qs ft})] 
(fluctuation theorem for the quantum-state statistics).
It contains the quantum work fluctuation theorem as a special case,
and if applied near equilibrium it reproduces
the out-of-time-order fluctuation-dissipation theorem \cite{TsujiShitaraUeda2018}, 
which connects chaotic properties of the system and a nonlinear response function.
We have discussed various aspects of the distribution function for the quantum-state statistics.
In particular, the negativity of the distribution is closely related to 
the quantum chaotic nature (non-integrability) of the underlying model Hamiltonian.
We have numerically demonstrated this for one-dimensional integrable and non-integrable quantum lattice models.
The implications of the obtained relations to thermodynamics and thermalization in isolated quantum systems
merit further study.

\acknowledgements

N.T. is supported by JSPS KAKENHI Grant No. JP16K17729.
M.U. acknowledges support by KAKENHI Grant No. JP26287088 and KAKENHI Grant No. JP15H05855.

\begin{widetext}

\appendix

\section{Proof of the fluctuation theorem for the quantum-state statistics}
\label{appendix A}

In this section, we prove the fluctuation theorem for the quantum-state statistics [Eq.~(\ref{qs ft})],
\begin{align}
\frac{p_n(w)}{\bar p_n(-w)}
&=
e^{\beta(w-n\Delta F(n\beta))}
\quad
(n=1,2,\dots).
\label{suppl:qs ft}
\end{align}
The proof is actually similar to that for the ordinary quantum work fluctuation theorem using
the method of characteristic functions \cite{Talkner2007}.

Let us first recursively evaluate the product $\circledast$ in the definition of $p_n(w)$
in the energy eigenbasis,
\begin{align}
p_n(w)
&=
\frac{1}{\mathcal N_n}{\rm Tr}([\![\hat\rho]\!]^{\circledast n}(w))
\notag
\\
&=
\frac{1}{\mathcal N_n}\sum_{k_1,\cdots,k_n} \sum_{l_1,\cdots,l_n}
p_{k_1}^i p_{k_2}^i \cdots p_{k_n}^i
\delta(w-(E_{l_1}^f+E_{l_2}^f+\cdots+E_{l_n}^f)+(E_{k_1}^i+E_{k_2}^i+\cdots+E_{k_n}^i))
\notag
\\
&\quad\times
\langle E_{l_1}^f|\hat U|E_{k_1}^i\rangle \langle E_{k_1}^i|\hat U^\dagger|E_{l_2}^f \rangle
\langle E_{l_2}^f|\hat U|E_{k_2}^i\rangle \langle E_{k_2}^i|\hat U^\dagger|E_{l_3}^f \rangle
\cdots
\langle E_{l_n}^f|\hat U|E_{k_n}^i\rangle \langle E_{k_n}^i|\hat U^\dagger|E_{l_1}^f \rangle.
\label{suppl:p_n}
\end{align}
The normalization constant $\mathcal N_n$ is determined by the direct calculation of the integral of $p_n(w)$,
\begin{align}
1
&=
\int_{-\infty}^{\infty} dw\, p_n(w)
\notag
\\
&=
\frac{1}{\mathcal N_n}\sum_{k_1,\cdots,k_n} \sum_{l_1,\cdots,l_n}
p_{k_1}^i p_{k_2}^i \cdots p_{k_n}^i
\langle E_{l_1}^f|\hat U|E_{k_1}^i\rangle \langle E_{k_1}^i|\hat U^\dagger|E_{l_2}^f \rangle
\langle E_{l_2}^f|\hat U|E_{k_2}^i\rangle \langle E_{k_2}^i|\hat U^\dagger|E_{l_3}^f \rangle
\cdots
\langle E_{l_n}^f|\hat U|E_{k_n}^i\rangle \langle E_{k_n}^i|\hat U^\dagger|E_{l_1}^f \rangle.
\notag
\\
&=
\frac{1}{\mathcal N_n}\sum_{l_1,\cdots,l_n}
\langle E_{l_1}^f|\hat U \hat\rho_i \hat U^\dagger|E_{l_2}^f \rangle
\langle E_{l_2}^f|\hat U \hat\rho_i \hat U^\dagger|E_{l_3}^f \rangle
\cdots
\langle E_{l_n}^f|\hat U \hat\rho_i \hat U^\dagger|E_{l_1}^f \rangle.
\notag
\\
&=
\frac{1}{\mathcal N_n}
{\rm Tr}[(\hat U \hat\rho_i \hat U^\dagger)
(\hat U \hat\rho_i \hat U^\dagger)
\cdots
(\hat U \hat\rho_i \hat U^\dagger)]
=
\frac{1}{\mathcal N_n}
{\rm Tr}(\hat\rho_i^n)
=
\frac{1}{\mathcal N_n}
\frac{Z_i(n\beta)}{Z_i(\beta)^n}.
\end{align}
Hence $\mathcal N_n$ is given by the equilibrium partition function as
\begin{align}
\mathcal N_n
&=
\frac{Z_i(n\beta)}{Z_i(\beta)^n}.
\label{suppl:N_n}
\end{align}
In particular, $\mathcal N_n$ is real ($\mathcal N_n\in\mathbb R$).
$p_n(w)$ is also real ($p_n(w)\in\mathbb R$) as confirmed by taking the complex conjugate of $p_n(w)$,
\begin{align}
p_n(w)^\ast
&=
\frac{1}{\mathcal N_n}\sum_{k_1,\cdots,k_n} \sum_{l_1,\cdots,l_n}
p_{k_1}^i p_{k_2}^i \cdots p_{k_n}^i
\delta(w-(E_{l_1}^f+E_{l_2}^f+\cdots+E_{l_n}^f)+(E_{k_1}^i+E_{k_2}^i+\cdots+E_{k_n}^i))
\notag
\\
&\quad\times
\langle E_{k_1}^i|\hat U^\dagger|E_{l_1}^f\rangle \langle E_{l_2}^f|\hat U|E_{k_1}^i \rangle
\langle E_{k_2}^i|\hat U^\dagger|E_{l_2}^f\rangle \langle E_{l_3}^f|\hat U|E_{k_2}^i \rangle
\cdots
\langle E_{k_n}^i|\hat U^\dagger|E_{l_n}^f\rangle \langle E_{l_1}^f|\hat U|E_{k_n}^i \rangle
\notag
\\
&=
\frac{1}{\mathcal N_n}\sum_{k_1,\cdots,k_n} \sum_{l_1,\cdots,l_n}
p_{k_1}^i p_{k_2}^i \cdots p_{k_n}^i
\delta(w-(E_{l_1}^f+E_{l_2}^f+\cdots+E_{l_n}^f)+(E_{k_1}^i+E_{k_2}^i+\cdots+E_{k_n}^i))
\notag
\\
&\quad\times
\langle E_{l_1}^f|\hat U|E_{k_n}^i \rangle \langle E_{k_n}^i|\hat U^\dagger|E_{l_n}^f\rangle
\cdots
\langle E_{l_3}^f|\hat U|E_{k_2}^i \rangle \langle E_{k_2}^i|\hat U^\dagger|E_{l_2}^f\rangle
\langle E_{l_2}^f|\hat U|E_{k_1}^i \rangle \langle E_{k_1}^i|\hat U^\dagger|E_{l_1}^f\rangle.
\label{suppl:p^ast}
\end{align}
By changing the summation labels as $k_i\to k_{n+1-i}$ and $l_i\to l_{n+1-i}$
and subsequently permuting the labels cyclicly, $l_1\to l_2\to\cdots\to l_n\to l_1$,
one can see that $p_n(w)^\ast$ (\ref{suppl:p^ast}) becomes identical to $p_n(w)$ (\ref{suppl:p_n}),
proving the realness of $p_n(w)$.

After Fourier transformation, the characteristic function $G_n(u)$ [Eq.~(\ref{G_n})] is given by
\begin{align}
G_n(u)
&=
\int_{-\infty}^{\infty} dw\, e^{iuw}p_n(w)
\notag
\\
&=
\frac{1}{\mathcal N_n}\sum_{k_1,\cdots,k_n,l_1,\cdots,l_n}
p_{k_1}^i \cdots p_{k_n}^i
e^{iu(E_{l_1}^f+\cdots+E_{l_n}^f)-iu(E_{k_1}^i+\cdots+E_{k_n}^i)}
\notag
\\
&\quad\times
\langle E_{l_1}^f|\hat U|E_{k_1}^i\rangle \langle E_{k_1}^i|\hat U^\dagger|E_{l_2}^f \rangle
\langle E_{l_2}^f|\hat U|E_{k_2}^i\rangle \langle E_{k_2}^i|\hat U^\dagger|E_{l_3}^f \rangle
\cdots
\langle E_{l_n}^f|\hat U|E_{k_n}^i\rangle \langle E_{k_n}^i|\hat U^\dagger|E_{l_1}^f \rangle.
\end{align}
Here we define an operator
\begin{align}
\hat W_{i,u}
&\equiv
e^{iu\hat H_i},
\\
\hat W_{f,u}
&\equiv
e^{iu\hat H_f}.
\end{align}
With this, the characteristic function can be expressed in a compact form of
\begin{align}
G_n(u)
&=
\frac{1}{\mathcal N_n}
{\rm Tr}[(\hat\rho_i \hat W_{i,u}^\dagger\hat U^\dagger\hat W_{f,u}\hat U)^n].
\label{suppl:G_n2}
\end{align}
If we take the Heisenberg picture, the explicit time dependence is included in the operators as
$\hat W_{i,u}^\dagger(t_i)=\hat W_{i,u}^\dagger$ and
$\hat W_{f,u}(t_f)=\hat U^\dagger\hat W_{f,u}\hat U^\dagger$,
with which $G_n(u)$ is written as
\begin{align}
G_n(u)
&=
\frac{1}{\mathcal N_n}
{\rm Tr}[(\hat\rho_i \hat W_{i,u}^\dagger(t_i)\hat W_{f,u}(t_f))^n].
\label{suppl:heisenberg picture}
\end{align}
One can see that for $n\ge 2$ the operators are out-of-time-ordered, i.e., Eq.~(\ref{suppl:heisenberg picture})
cannot be expressed as the usual time-ordered product. Hence $G_n(u)$ ($n\ge 2$) is classified as an out-of-time-ordered
correlator.

If we expand the trace in Eq.~(\ref{suppl:G_n2}) in a complete basis set $\{|m\rangle \}_m$, $G_n(u)$ is written as
\begin{align}
G_n(u)
&=
\frac{1}{\mathcal N_n}
\sum_m \langle m|
(\hat\rho_i \hat W_{i,u}^\dagger\hat U^\dagger\hat W_{f,u}\hat U)^n
|m\rangle.
\end{align}
Here we use the identity $\langle k|\hat O|l\rangle=\langle\bar k|\Theta\hat O^\dagger\Theta^{-1}|\bar l\rangle$ \cite{SakuraiBook}
which is valid
for arbitrary linear operators $\hat O$, where $\Theta$ is the antiunitary time-reversal operator,
$|\bar k\rangle\equiv\Theta|k\rangle$ and $|\bar l\rangle\equiv\Theta|l\rangle$,
to obtain
\begin{align}
G_n(u)
&=
\frac{1}{\mathcal N_n}
\sum_m \langle \bar m|\Theta
(\hat U^\dagger \hat W_{f,u}^\dagger \hat U \hat W_{i,u}\hat\rho_i)^n
\Theta^{-1}|\bar m\rangle.
\end{align}
We define the time-reversed counterpart of the operators $\hat W_{i,u}$ and $\hat W_{f,u}$,
\begin{align}
\hat{\bar W}_{i,u}
&\equiv
e^{iu\hat{\bar H}_i},
\\
\hat{\bar W}_{f,u}
&\equiv
e^{iu\hat{\bar H}_f}.
\end{align}
Let us recall that $\Theta\hat U^\dagger\Theta^{-1}=\hat{\bar U}$, 
$\Theta\hat W_{f,u}^\dagger\Theta^{-1}=\hat{\bar W}_{i,u}$,
$\Theta\hat U\Theta^{-1}=\hat{\bar U}^\dagger$,
$\Theta\hat W_{i,u}\Theta^{-1}=\hat{\bar W}_{f,u}$, and
$\Theta\hat\rho_i\Theta^{-1}=\bar Z_f(\beta) Z_i(\beta)^{-1}\hat{\bar\rho}_f$
(since $\Theta\hat H_i\Theta^{-1}=\hat{\bar H}_f$). From these, we have
\begin{align}
G_n(u)
&=
\frac{1}{\mathcal N_n}
\frac{\bar Z_f(\beta)^n}{Z_i(\beta)^n}
{\rm Tr}[
(\hat{\bar U} \hat{\bar W}_{i,u} \hat{\bar U}^\dagger \hat{\bar W}_{f,u}^\dagger \hat{\bar\rho}_f)^n].
\end{align}
Now we use the following relations,
\begin{align}
\hat{\bar W}_{f,u}^\dagger\hat{\bar\rho}_f
&=
\bar Z_f(\beta)^{-1}\hat{\bar W}_{f,u-i\beta}^\dagger,
\\
\hat{\bar W}_{i,u}
&=
\bar Z_i(\beta)\hat{\bar\rho}_i\hat{\bar W}_{i,u-i\beta}.
\end{align}
Then $G_n(u)$ is written as
\begin{align}
G_n(u)
&=
\frac{1}{\mathcal N_n}
\frac{\bar Z_i(\beta)^n}{Z_i(\beta)^n}
{\rm Tr}[
(\hat{\bar U} \hat{\bar\rho}_i\hat{\bar W}_{i,u-i\beta} \hat{\bar U}^\dagger \hat{\bar W}_{f,u-i\beta}^\dagger)^n]
\notag
\\
&=
\frac{1}{\mathcal N_n}
\frac{\bar Z_i(\beta)^n}{Z_i(\beta)^n}
{\rm Tr}[
(\hat{\bar\rho}_i\hat{\bar W}_{i,u-i\beta} \hat{\bar U}^\dagger \hat{\bar W}_{f,u-i\beta}^\dagger\hat{\bar U})^n],
\end{align}
where we performed the cyclic permutation in the trace. We further rewrite $G_n(u)$ using the relations,
\begin{align}
\hat{\bar W}_{i,u-i\beta}
&=
\hat{\bar W}_{i,-u+i\beta}^\dagger,
\\
\hat{\bar W}_{f,u-i\beta}^\dagger
&=
\hat{\bar W}_{f,-u+i\beta}.
\end{align}
They lead to
\begin{align}
G_n(u)
&=
\frac{1}{\mathcal N_n}
\frac{\bar Z_i(\beta)^n}{Z_i(\beta)^n}
{\rm Tr}[
(\hat{\bar\rho}_i\hat{\bar W}_{i,-u+i\beta}^\dagger \hat{\bar U}^\dagger \hat{\bar W}_{f,-u+i\beta}\hat{\bar U})^n].
\label{suppl:G_n}
\end{align}
One can notice that
the right-hand side of Eq.~(\ref{suppl:G_n}) is proportional to the characteristic function for $\bar p_n(w)$,
\begin{align}
\bar G_n(u)
&\equiv
\int_{-\infty}^{\infty} dw\, e^{iuw}\bar p_n(w)
\notag
\\
&=
\frac{1}{\bar{\mathcal N}_n}
{\rm Tr}[(\hat{\bar\rho}_i \hat{\bar W}_{i,u}^\dagger\hat{\bar U}^\dagger\hat{\bar W}_{f,u}\hat{\bar U})^n]
\label{suppl:G_n bar}
\end{align}
In the same way as for $\mathcal N_n$,
the normalization constant $\bar{\mathcal N}_n$ is given by
\begin{align}
\bar{\mathcal N}_n
&=
\frac{\bar Z_i(n\beta)}{\bar Z_i(\beta)^n}.
\label{suppl:N_n bar}
\end{align}
The partition function for the time-reversed process is related to the one for the forward process through
\begin{align}
\bar Z_i(\beta)
&=
{\rm Tr}(e^{-\beta\hat{\bar H}_i})
=
{\rm Tr}(e^{-\beta\Theta\hat H_f\Theta^{-1}})
=
{\rm Tr}(\Theta e^{-\beta\hat H_f}\Theta^{-1})
=
{\rm Tr}(e^{-\beta\hat H_f})
=
Z_f(\beta).
\label{suppl:Z_i bar}
\end{align}
By comparing Eq.~(\ref{suppl:G_n}) with Eq.~(\ref{suppl:G_n bar}) and using Eqs.~(\ref{suppl:N_n}), (\ref{suppl:N_n bar}), and (\ref{suppl:Z_i bar}),
we arrive at the symmetry relation,
\begin{align}
G_n(u)
&=
\frac{Z_f(n\beta)}{Z_i(n\beta)}\bar G_n(-u+i\beta).
\end{align}
Its inverse Fourier transformation gives
\begin{align}
p_n(w)
&=
\frac{Z_f(n\beta)}{Z_i(n\beta)} e^{\beta w}\bar p_n(-w).
\end{align}
Finally, the partition function can be expressed in terms of the equilibrium free energy, 
$Z_{i,f}(n\beta)=e^{-n\beta F_{i,f}(n\beta)}$,
with which the fluctuation theorem for the quantum-state statistics (\ref{suppl:qs ft}) is proved.
\hspace{\fill}$\blacksquare$

\section{Derivation of the out-of-time-order fluctuation-dissipation theorem from the fluctuation theorem for quantum-state statistics}
\label{appendix B}

In this section, we show the derivation of the out-of-time-order fluctuation-dissipation theorem (FDT) \cite{TsujiShitaraUeda2018}
around zero frequency from the fluctuation theorem for the quantum-state statistics [Eq.~(\ref{qs ft})]. 

Before looking into the details of the derivation, 
let us overview the derivation of the ordinary fluctuation-dissipation theorem around zero frequency 
from the quantum work fluctuation theorem [Eq.~(\ref{qs ft}) with $n=1$], 
which helps one to understand the derivation of the out-of-time-order version.
Here we mean the fluctuation-dissipation theorem in the form of \cite{Kubo1957,TsujiShitaraUeda2018}
\begin{align}
C_{\{A,B\}}(\omega)
&=
\coth\left(\frac{\beta\hbar\omega}{2}\right)
C_{[A,B]}(\omega),
\label{suppl:FDT}
\end{align}
where $\hat A$ and $\hat B$ are arbitrary observables, and
$C_{\{A,B\}}(\omega)$ and $C_{[A,B]}(\omega)$ are Fourier transforms of the anticommutator and commutator
correlation functions, respectively,
\begin{align}
C_{\{A,B\}}(\omega)
&\equiv
\int_{-\infty}^{\infty} dt\, e^{i\omega t} \langle \{\hat A(t), \hat B(0)\}\rangle,
\\
C_{[A,B]}(\omega)
&\equiv
\int_{-\infty}^{\infty} dt\, e^{i\omega t} \langle [\hat A(t), \hat B(0)]\rangle,
\end{align}
where $\langle \cdots \rangle\equiv{\rm Tr}[\hat\rho_i \cdots ]$ denotes the statistical average over the initial state.
To derive the FDT, we perform the cumulant expansion of the integrated fluctuation theorem
$\langle e^{-\beta w}\rangle_{p_1}=e^{-\beta \Delta F(\beta)}$ (Jarzynski equality)
up to the second order,
\begin{align}
\langle w\rangle_{p_1}-\Delta F(\beta)
&\approx
\frac{\beta}{2}\langle (\Delta w)^2\rangle_{p_1},
\label{suppl:Jarzynski2}
\end{align}
with $\Delta w\equiv w-\langle w\rangle_{p_1}$. 
The approximation ($\approx$) means that we have neglected the $k$th-order cumulant terms for $k\ge 3$.
The cumulant expansion in (\ref{suppl:Jarzynski2}) corresponds to the expansion of (\ref{suppl:FDT})
around zero frequency.
To evaluate $\langle w\rangle_p$ and $\langle (\Delta w)^2\rangle_p$, we use the characteristic function for $p_1(w)$,
\begin{align}
G_1(u)
&=
\int_{-\infty}^{\infty} dw\, e^{iuw}p_1(w)
=
\langle \hat W_i^\dagger(u)\hat U^\dagger\hat W_f(u)\hat U\rangle.
\end{align}
By taking the derivatives of $G_1(u)$, we obtain
\begin{align}
\langle w\rangle_{p_1}
&=
\frac{\partial G_1(u)}{\partial iu}\bigg|_{u=0}
=
\langle \hat U^\dagger\hat H_f \hat U\rangle
-\langle \hat H_i \rangle,
\label{suppl:w}
\\
\langle (\Delta w)^2\rangle_{p_1}
&=
\frac{\partial^2 G_1(u)}{\partial (iu)^2}\bigg|_{u=0}
-\langle w\rangle_{p_1}^2
=
\langle \hat U^\dagger\hat H_f^2 \hat U\rangle
-2\langle \hat H_i \hat U^\dagger \hat H_f \hat U\rangle
+\langle \hat H_i^2\rangle-\langle w\rangle_{p_1}^2.
\label{suppl:w2}
\end{align}

The fluctuation theorem is valid for arbitrary perturbations. Here we consider a specific form
of the perturbation,
\begin{align}
\hat H(s)
&=
\hat H_0+\xi(s)\hat X_S(s),
\quad
(t_i\le s\le t_f)
\label{suppl:perturbation}
\end{align}
where $\hat H_0$ is the time-independent unperturbed Hamiltonian,
$\hat X_S(s)$ represents the external force in the Schr\"odinger picture, and $\xi(s)$ is a time-dependent parameter
($\xi(s)\in\mathbb R$).
In the Heisenberg picture, we denote $\hat X(s)=\hat U(s,t_i)^\dagger\hat X_S(s)\hat U(s,t_i)$
[$\hat U(t,t')\equiv\mathcal T\exp(-\frac{i}{\hbar}\int_{t'}^t ds\, \hat H(s))$ for $t\ge t'$].
In order for $\hat H(s)$ to be hermitian, $\hat X(s)$ should also be hermitian.
Suppose that, after the system is driven by the external force, the Hamiltonian comes back to the initial one
($\hat H_i=\hat H_f=\hat H_0$). In this case, the free-energy difference vanishes ($\Delta F(\beta)=0$).
By taking the second functional derivative with respect to $\xi(s)$ on both sides of Eq.~(\ref{suppl:Jarzynski2})
and putting $\xi(s)=0$, we obtain
\begin{align}
\frac{\delta^2}{\delta\xi(t_1)\delta\xi(t_2)}\langle w\rangle_{p_1}\bigg|_{\xi=0}
&\approx
\frac{\beta}{2}\frac{\delta^2}{\delta\xi(t_1)\delta\xi(t_2)}\langle (\Delta w)^2\rangle_{p_1}\bigg|_{\xi=0}
\label{suppl:derivative2}
\end{align}
with $t_i<t_2\le t_1<t_f$. Using Eq.~(\ref{suppl:w}), the left-hand side of Eq.~(\ref{suppl:derivative2}) is calculated as
\begin{align}
\frac{\delta^2}{\delta\xi(t_1)\delta\xi(t_2)}\langle w\rangle_{p_1}\bigg|_{\xi=0}
&=
\left(\frac{i}{\hbar}\right)^2\langle [\hat X(t_2), [\hat X(t_1), \hat H_0]]\rangle
=
\frac{i}{\hbar}\langle [\dot{\hat X}(t_1), \hat X(t_2)] \rangle,
\end{align}
where $\dot{\hat X}(s)\equiv \frac{i}{\hbar}[\hat H_0, \hat X(s)]$ does not include a derivative in terms of
the explicit time dependence of $\hat X_S(s)$. Using Eq.~(\ref{suppl:w2}), the right-hand side of Eq.~(\ref{suppl:derivative2})
is calculated as
\begin{align}
\frac{\delta^2}{\delta\xi(t_1)\delta\xi(t_2)}\langle (\Delta w)^2\rangle_{p_1}\bigg|_{\xi=0}
&=
\left(\frac{i}{\hbar}\right)^2\langle [\hat X(t_2), [\hat X(t_1), \hat H_0^2]]\rangle
-2\left(\frac{i}{\hbar}\right)^2\langle \hat H_0 [\hat X(t_2), [\hat X(t_1), \hat H_0]]\rangle
\notag
\\
&=
\langle \{\dot{\hat X}(t_1), \dot{\hat X}(t_2)\}\rangle
=
-\langle \{\ddot{\hat X}(t_1), \hat X(t_2)\}\rangle.
\end{align}
By substituting these results in Eq.~(\ref{suppl:derivative2}), we obtain
\begin{align}
\langle [\dot{\hat X}(t_1), \hat X(t_2)] \rangle
&\approx
\frac{i\beta\hbar}{2}\langle \{\ddot{\hat X}(t_1), \hat X(t_2)\}\rangle.
\label{suppl:FDT X}
\end{align}
So far, we have assumed $t_1\ge t_2$. However, the relation (\ref{suppl:FDT X}) also holds for $t_1< t_2$, which can be confirmed
by exchanging $t_1$ and $t_2$ in Eq.~(\ref{suppl:FDT X}). By taking the limit of $t_i\to -\infty$ and $t_f\to\infty$, 
one can see that the relation (\ref{suppl:FDT X}) is valid for arbitrary $t_1$ and $t_2$.
If we write $\dot{\hat X}(t_1)=:\hat A(t_1)$ and $\hat X(t_2)=:\hat B(t_2)$ ($\hat A$ and $\hat B$ are arbitrary hermitian operators), we have
\begin{align}
\langle [\hat A(t_1), \hat B(t_2)] \rangle
&\approx
\frac{\beta\hbar}{2}i\partial_{t_1}\langle \{\hat A(t_1), \hat B(t_2)\}\rangle
\label{suppl:Jarzynski derived}
\\
\Leftrightarrow\quad
C_{[A,B]}(\omega)
&\approx
\frac{\beta\hbar\omega}{2}C_{\{A,B\}}(\omega).
\end{align}
This is nothing but the near-zero-frequency part ($\omega\sim 0$) of the ordinary FDT (\ref{suppl:FDT}).

Now, we move on to the derivation of the out-of-time-order FDT \cite{TsujiShitaraUeda2018},
which can be expressed in the form of
\begin{align}
C_{\{A,B\}^2}(\omega)+C_{[A,B]^2}(\omega)
&=
2\coth\left(\frac{\beta\hbar\omega}{4}\right)
C_{\{A,B\}[A,B]}(\omega),
\label{suppl:OTO FDT}
\end{align}
where we have defined
\begin{align}
C_{\{A,B\}^2}(\omega)
&\equiv
\int_{-\infty}^{\infty} dt\, e^{i\omega t}
\langle \{\hat A(t), \hat B(0)\}, \{\hat A(t), \hat B(0)\}\rangle,
\\
C_{[A,B]^2}(\omega)
&\equiv
\int_{-\infty}^{\infty} dt\, e^{i\omega t}
\langle [\hat A(t), \hat B(0)], [\hat A(t), \hat B(0)]\rangle,
\\
C_{\{A,B\}[A,B]}(\omega)
&\equiv
\int_{-\infty}^{\infty} dt\, e^{i\omega t}
\langle \{\hat A(t), \hat B(0)\}, [\hat A(t), \hat B(0)]\rangle.
\end{align}
Here we use the notation of the bipartite statistical average with respect to the initial state, 
$\langle \hat X, \hat Y\rangle={\rm Tr}(\hat\rho_i^{\frac{1}{2}}\hat X\hat\rho_i^{\frac{1}{2}}\hat Y)$.
The out-of-time-order FDT (\ref{suppl:OTO FDT}) can be derived from 
the cumulant expansion of the integrated
form of the fluctuation theorem for the quantum-state statistics [Eq.~(\ref{qs Jarzynski}) with $n=2$] at temperature $(\beta/2)^{-1}$
together with the ordinary integrated fluctuation theorem [Eq.~(\ref{qs Jarzynski}) with $n=1$] at temperature $\beta^{-1}$
up to the third order,
\begin{align}
\langle w\rangle_{p_1,\beta}-\Delta F(\beta)
&\approx
\frac{\beta}{2}\langle (\Delta w)^2\rangle_{p_1,\beta}
-\frac{\beta^2}{6}\langle (\Delta w)^3\rangle_{p_1,\beta},
\label{suppl:Jarzynski3}
\\
\frac{1}{2}\langle w\rangle_{p_2,\frac{\beta}{2}}-\Delta F(\beta)
&\approx
\frac{\beta}{8}\langle (\Delta w)^2\rangle_{p_2,\frac{\beta}{2}}
-\frac{\beta^2}{48}\langle (\Delta w)^3\rangle_{p_2,\frac{\beta}{2}}.
\label{suppl:oto Jarzynski3}
\end{align}
Here we explicitly show the temperature at which the expectation value is evaluated.
As we stated for the ordinary FDT (\ref{suppl:FDT}),
the cumulant expansion here corresponds to the expansion of (\ref{suppl:OTO FDT}) around zero frequency.
Hereafter we focus on the leading terms around zero frequency, and neglect higher-order derivatives in time
during the derivation.
We subtract both sides of Eq.~(\ref{suppl:Jarzynski3}) from those of Eq.~(\ref{suppl:oto Jarzynski3}),
\begin{align}
\frac{1}{2}
\langle w\rangle_{p_2,\frac{\beta}{2}}-\langle w\rangle_{p_1,\beta}
&\approx
\frac{\beta}{2}\left[
\frac{1}{4}\langle (\Delta w)^2\rangle_{p_2,\frac{\beta}{2}}-\langle (\Delta w)^2\rangle_{p_1,\beta}
\right]
-\frac{\beta^2}{6}\left[
\frac{1}{8}\langle (\Delta w)^3\rangle_{p_2,\frac{\beta}{2}}-\langle (\Delta w)^3\rangle_{p_1,\beta}
\right].
\label{suppl:Jarzynski diff}
\end{align}
The explicit forms of $\langle w\rangle_{p_1,\beta}$ and $\langle (\Delta w)^2\rangle_{p_1,\beta}$ are given in Eqs.~(\ref{suppl:w}) and (\ref{suppl:w2}), respectively.
$\langle (\Delta w)^3\rangle_{p_1,\beta}$ is given by
\begin{align}
\langle (\Delta w)^3\rangle_{p_1,\beta}
&=
\frac{\partial^3 G_1(u)}{\partial(iu)^3}\bigg|_{u=0}
-3\langle w^2\rangle_{p_1,\beta} \langle w\rangle_{p_1,\beta}
+2\langle w\rangle_{p_1,\beta}^3
\notag
\\
&=
\langle \hat U^\dagger\hat H_f^3 \hat U\rangle
-3\langle \hat H_i\hat U^\dagger\hat H_f^2 \hat U\rangle
+3\langle \hat H_i^2\hat U^\dagger\hat H_f \hat U\rangle
-\langle \hat H_i^3\rangle
-3\langle w^2\rangle_{p_1,\beta} \langle w\rangle_{p_1,\beta}
+2\langle w\rangle_{p_1,\beta}^3.
\end{align}
To evaluate the remaining $\langle w\rangle_{p_2,\frac{\beta}{2}}$, $\langle (\Delta w)^2\rangle_{p_2,\frac{\beta}{2}}$,
and $\langle (\Delta w)^3\rangle_{p_2,\frac{\beta}{2}}$ in Eq.~(\ref{suppl:Jarzynski diff}), 
we use the characteristic function for $p_2(w)$,
\begin{align}
G_2(u)|_{\frac{\beta}{2}}
&\equiv
\int_{-\infty}^{\infty} dw\, e^{iuw}p_2(w)|_{\frac{\beta}{2}}
=
\langle \hat W_i^\dagger(u)\hat U^\dagger\hat W_f(u)\hat U,
\hat W_i^\dagger(u)\hat U^\dagger\hat W_f(u)\hat U\rangle.
\end{align}
$\langle w\rangle_{p_2,\frac{\beta}{2}}$, $\langle (\Delta w)^2\rangle_{p_2,\frac{\beta}{2}}$,
and $\langle (\Delta w)^3\rangle_{p_2,\frac{\beta}{2}}$ are provided by the derivatives of $G_2(u)|_{\frac{\beta}{2}}$,
\begin{align}
\langle w\rangle_{p_2,\frac{\beta}{2}}
&=
\frac{\partial G_2(u)}{\partial (iu)}\bigg|_{\frac{\beta}{2},u=0}
=
2\langle \hat U^\dagger\hat H_f\hat U\rangle-2\langle \hat H_i\rangle,
\\
\langle (\Delta w)^2\rangle_{p_2,\frac{\beta}{2}}
&=
\frac{\partial^2 G_2(u)}{\partial (iu)^2}\bigg|_{\frac{\beta}{2},u=0}
-\langle w\rangle_{p_2,\frac{\beta}{2}}^2
\notag
\\
&=
2\langle \hat U^\dagger \hat H_f^2 \hat U\rangle
+2\langle \hat U^\dagger\hat H_f \hat U, \hat U^\dagger\hat H_f\hat U\rangle
+4\langle \hat H_i^2\rangle
-8\langle \hat H_i \hat U^\dagger\hat H_f \hat U\rangle
-\langle w\rangle_{p_2,\frac{\beta}{2}}^2,
\\
\langle (\Delta w)^3\rangle_{p_2,\frac{\beta}{2}}
&=
\frac{\partial^3 G_2(u)}{\partial(iu)^3}\bigg|_{\frac{\beta}{2},u=0}
-3\langle w^2\rangle_{p_2,\frac{\beta}{2}} \langle w\rangle_{p_2,\frac{\beta}{2}}
+2\langle w\rangle_{p_2,\frac{\beta}{2}}^3
\notag
\\
&=
2\langle \hat U^\dagger\hat H_f^3 \hat U\rangle
+6\langle \hat U^\dagger\hat H_f^2 \hat U, \hat U^\dagger\hat H_f \hat U\rangle
-12\langle \hat H_i\hat U^\dagger\hat H_f^2 \hat U\rangle
\notag
\\
&\quad
-12\langle \hat H_i\hat U^\dagger\hat H_f \hat U, \hat U^\dagger\hat H_f \hat U\rangle
+24\langle \hat H_i^2\hat U^\dagger\hat H_f \hat U\rangle
-8\langle \hat H_i^3\rangle
-3\langle w^2\rangle_{p_2,\frac{\beta}{2}} \langle w\rangle_{p_2,\frac{\beta}{2}}
+2\langle w\rangle_{p_2,\frac{\beta}{2}}^3.
\end{align}
By subtracting $\langle w\rangle_{p1,\beta}$, $\langle (\Delta w)^2\rangle_{p_1,\beta}$, and $\langle (\Delta w)^3\rangle_{p_1,\beta}$
from $\frac{1}{2}\langle w\rangle_{p_2,\frac{\beta}{2}}$, $\frac{1}{4}\langle (\Delta w)^2\rangle_{p_2,\frac{\beta}{2}}$, and $\frac{1}{8}\langle (\Delta w)^3\rangle_{p_2,\frac{\beta}{2}}$, respectively, we obtain
\begin{align}
\frac{1}{2}\langle w\rangle_{p_2,\frac{\beta}{2}}-\langle w\rangle_{p_1,\beta}
&=
0,
\\
\frac{1}{4}\langle (\Delta w)^2\rangle_{p_2,\frac{\beta}{2}}-\langle (\Delta w)^2\rangle_{p_1,\beta}
&=
-\frac{1}{2}\langle \hat U^\dagger \hat H_f^2 \hat U\rangle
+\frac{1}{2}\langle \hat U^\dagger\hat H_f \hat U, \hat U^\dagger\hat H_f\hat U\rangle,
\\
\frac{1}{8}\langle (\Delta w)^3\rangle_{p_2,\frac{\beta}{2}}-\langle (\Delta w)^3\rangle_{p_1,\beta}
&=
-\frac{3}{4}\langle \hat U^\dagger\hat H_f^3 \hat U\rangle
+\frac{3}{4}\langle \hat U^\dagger\hat H_f^2 \hat U, \hat U^\dagger\hat H_f \hat U\rangle
+\frac{3}{2}\langle \hat H_i\hat U^\dagger\hat H_f^2 \hat U\rangle
\notag
\\
&\quad
-\frac{3}{2}\langle \hat H_i\hat U^\dagger\hat H_f \hat U, \hat U^\dagger\hat H_f \hat U\rangle
-3\langle w\rangle_{p_1,\beta}
\left[\frac{1}{4}\langle (\Delta w)^2\rangle_{p_2,\frac{\beta}{2}}-\langle (\Delta w)^2\rangle_{p_1,\beta}\right].
\end{align}

As in the case of the ordinary fluctuation theorem, we consider a perturbation in the form of (\ref{suppl:perturbation}).
We take the second derivative of both sides of Eq.~(\ref{suppl:Jarzynski diff}) with respect to $\xi(s)$
and put $\xi(s)=0$. This results in
\begin{align}
\frac{\delta^2}{\delta\xi(t_1)\delta\xi(t_2)}\left[\frac{1}{4}\langle (\Delta w)^2\rangle_{p_2,\frac{\beta}{2}}-\langle (\Delta w)^2\rangle_{p_1,\beta}\right]\bigg|_{\xi=0}
&\approx
\frac{\beta}{3}\frac{\delta^2}{\delta\xi(t_1)\delta\xi(t_2)}
\left[\frac{1}{8}\langle (\Delta w)^3\rangle_{p_2,\frac{\beta}{2}}-\langle (\Delta w)^3\rangle_{p_1,\beta}\right]\bigg|_{\xi=0}.
\label{suppl:Jarzynski difference second derivative}
\end{align}
The left hand side of Eq.~(\ref{suppl:Jarzynski difference second derivative}) is calculated as
\begin{align}
&
\frac{\delta^2}{\delta\xi(t_1)\delta\xi(t_2)}\left[\frac{1}{4}\langle (\Delta w)^2\rangle_{p_2,\frac{\beta}{2}}-\langle (\Delta w)^2\rangle_{p_1,\beta}\right]\bigg|_{\xi=0}
\notag
\\
&=
-\frac{1}{2}\Big(\frac{i}{\hbar}\Big)^2 \langle [\hat X(t_2), [\hat X(t_1), \hat H_0^2]] \rangle
+\Big(\frac{i}{\hbar}\Big)^2 \langle \hat H_0 [\hat X(t_2), [\hat X(t_1), \hat H_0]] \rangle
+\Big(\frac{i}{\hbar}\Big)^2 \langle [\hat X(t_1), \hat H_0], [\hat X(t_2), \hat H_0] \rangle
\notag
\\
&=
-\frac{1}{2}\langle \{\dot{\hat X}(t_1), \dot{\hat X}(t_2)\} \rangle
+\langle \dot{\hat X}(t_1), \dot{\hat X}(t_2) \rangle
\end{align}
The second derivative in the right hand side of Eq.~(\ref{suppl:Jarzynski difference second derivative}) is calculated as
\begin{align}
&
\frac{\delta^2}{\delta\xi(t_1)\delta\xi(t_2)}
\left[\frac{1}{8}\langle (\Delta w)^3\rangle_{p_2,\frac{\beta}{2}}-\langle (\Delta w)^3\rangle_{p_1,\beta}\right]\bigg|_{\xi=0}
\notag
\\
&=
-\frac{3}{4}\Big(\frac{i}{\hbar}\Big)^2\langle [\hat X(t_2), [\hat X(t_1), \hat H_0^3]] \rangle
+\frac{3}{4}\Big(\frac{i}{\hbar}\Big)^2\langle [\hat X(t_2), [\hat X(t_1), \hat H_0^2]]\hat H_0\rangle
+\frac{3}{4}\Big(\frac{i}{\hbar}\Big)^2\langle \hat H_0^2[\hat X(t_2), [\hat X(t_1), \hat H_0]]\rangle
\notag
\\
&\quad
+\frac{3}{4}\Big(\frac{i}{\hbar}\Big)^2\langle [\hat X(t_1), \hat H_0^2], [\hat X(t_2), \hat H_0]\rangle
+\frac{3}{4}\Big(\frac{i}{\hbar}\Big)^2\langle [\hat X(t_2), \hat H_0^2], [\hat X(t_1), \hat H_0]\rangle
+\frac{3}{2}\Big(\frac{i}{\hbar}\Big)^2\langle \hat H_0[\hat X(t_2), [\hat X(t_1), \hat H_0^2]]\rangle
\notag
\\
&=
-i\hbar\frac{3}{4}\langle [\ddot{\hat X}(t_1), \dot{\hat X}(t_2)]\rangle.
\end{align}
Combining these results, we obtain from Eq.~(\ref{suppl:Jarzynski difference second derivative})
\begin{align}
\langle \dot{\hat X}(t_1), \dot{\hat X}(t_2) \rangle
&\approx
\frac{1}{2}\langle \{\dot{\hat X}(t_1), \dot{\hat X}(t_2)\} \rangle
-\frac{i\beta\hbar}{4}\langle [\ddot{\hat X}(t_1), \dot{\hat X}(t_2)]\rangle.
\label{suppl:OTO FDT X}
\end{align}
The relation (\ref{suppl:OTO FDT X}) can be viewed as the leading gradient expansion of
\begin{align}
\langle \dot{\hat X}(t_1), \dot{\hat X}(t_2) \rangle
&=
\frac{1}{2}\langle \dot{\hat X}(t_1-\frac{i\beta\hbar}{2}) \dot{\hat X}(t_2)\rangle
+\frac{1}{2}\langle \dot{\hat X}(t_2)\dot{\hat X}(t_1+\frac{i\beta\hbar}{2})\rangle.
\end{align}
As we will see below, this is almost equivalent to the out-of-time-order FDT (\ref{suppl:OTO FDT}).
Using the relation (\ref{suppl:FDT X}) and neglecting higher-order derivatives, one can also write
\begin{align}
\langle \dot{\hat X}(t_1), \dot{\hat X}(t_2) \rangle
&\approx
\frac{1}{2}\langle \{\dot{\hat X}(t_1), \dot{\hat X}(t_2)\} \rangle.
\label{suppl:OTO FDT X2}
\end{align}
We note that the relations (\ref{suppl:FDT X}), (\ref{suppl:OTO FDT X}) and (\ref{suppl:OTO FDT X2}) hold not only for hermitian operators $\hat X(t_1)$ and $\hat X(t_2)$
but also for arbitrary linear operators $\hat X(t_1)$ and $\hat X(t_2)$.
This is because we can decompose arbitrary operators $\hat X(t_1)$ and $\hat X(t_2)$
into a linear combination of hermitian terms
$\hat X(t_j)=\frac{1}{2}(\hat X(t_j)+\hat X(t_j)^\dagger)+\frac{1}{2i} i(\hat X(t_j)-\hat X(t_j)^\dagger)$
($j=1,2$)
and for each hermitian term we can apply (\ref{suppl:FDT X}), (\ref{suppl:OTO FDT X}) and (\ref{suppl:OTO FDT X2}).

The relation (\ref{suppl:OTO FDT X2}) together with (\ref{suppl:FDT X}) contains
enough information to reproduce the out-of-time-order FDT (\ref{suppl:OTO FDT}).
By substituting $\dot{\hat X}(t_1)=\{\hat A(t), \hat B(t')\}$ and $\dot{\hat X}(t_2)=[\hat A(t), \hat B(t')]$ in (\ref{suppl:OTO FDT X2}),
the right-hand side of (\ref{suppl:OTO FDT}) is approximated as
\begin{align}
\langle \{\hat A(t), \hat B(t')\}, [\hat A(t), \hat B(t')]\rangle
&\approx
\frac{1}{2}\langle \{\{\hat A(t), \hat B(t')\}, [\hat A(t), \hat B(t')]\}\rangle
=
\langle [\hat A(t), \hat B(t')\hat A(t)\hat B(t')] \rangle.
\label{suppl:OTO FDT right}
\end{align}
We then use (\ref{suppl:FDT X}) with $\dot X(t_1)=\hat A(t)$ and $\hat X(t_2)=\hat B(t')\hat A(t)\hat B(t')$ to have
\begin{align}
\langle [\hat A(t), \hat B(t')\hat A(t)\hat B(t')] \rangle
&\approx
\frac{i\beta\hbar}{2}
\langle \{\dot{\hat A}(t), \hat B(t')\hat A(t)\hat B(t')\} \rangle
\notag
\\
&=
\frac{i\beta\hbar}{4}\langle \{\dot{\hat A}(t)\hat B(t'), \hat A(t)\hat B(t')\}\rangle
+\frac{i\beta\hbar}{4}\langle [\dot{\hat A}(t)\hat B(t'), \hat A(t)\hat B(t')]\rangle
\notag
\\
&\quad
+\frac{i\beta\hbar}{4}\langle \{\hat B(t')\hat A(t), \hat B(t')\dot{\hat A}(t)\}\rangle
+\frac{i\beta\hbar}{4}\langle [\hat B(t')\hat A(t), \hat B(t')\dot{\hat A}(t)]\rangle.
\label{suppl:[A,BAB]}
\end{align}
From (\ref{suppl:FDT X}), one can see that the terms including commutators in Eq.~(\ref{suppl:[A,BAB]}) have higher-order derivatives,
which can be neglected here. The anticommutator terms in Eq.~(\ref{suppl:[A,BAB]}) can be expressed in terms of the bipartite
statistical average via (\ref{suppl:OTO FDT X2}),
\begin{align}
\langle [\hat A(t), \hat B(t')\hat A(t)\hat B(t')] \rangle
&\approx
\frac{i\beta\hbar}{2}\langle \dot{\hat A}(t)\hat B(t'), \hat A(t)\hat B(t')\rangle
+\frac{i\beta\hbar}{2}\langle \hat B(t')\hat A(t), \hat B(t')\dot{\hat A}(t)\rangle
\notag
\\
&=
\frac{\beta\hbar}{8}i\partial_t
\left(
\langle \{\hat A(t), \hat B(t')\}, \{\hat A(t), \hat B(t')\}\rangle
+\langle [\hat A(t), \hat B(t')], [\hat A(t), \hat B(t')]\rangle
\right).
\label{suppl:OTO FDT left}
\end{align}
Combining (\ref{suppl:OTO FDT right}), (\ref{suppl:[A,BAB]}), and (\ref{suppl:OTO FDT left}),
one can reproduce the near-zero-frequency part of the out-of-time-order FDT (\ref{suppl:OTO FDT}):
\begin{align}
\langle \{\hat A(t), \hat B(t')\}, [\hat A(t), \hat B(t')]\rangle
&\approx
\frac{\beta\hbar}{8}i\partial_t
\left(
\langle \{\hat A(t), \hat B(t')\}, \{\hat A(t), \hat B(t')\}\rangle
+\langle [\hat A(t), \hat B(t')], [\hat A(t), \hat B(t')]\rangle
\right)
\\
\Leftrightarrow\quad
C_{\{A,B\}[A,B]}(\omega)
&\approx
\frac{\beta\hbar\omega}{8} [C_{\{A,B\}^2}(\omega)+C_{[A,B]^2}(\omega)].
\label{suppl:OTO FDT approx}
\end{align}
Note that $[2\coth\big(\frac{\beta\hbar\omega}{4}\big)]^{-1}
=\frac{\beta\hbar\omega}{8}+O((\beta\hbar\omega)^2)$. 

\section{Numerical calculation of the distribution function for the quantum-state statistics}
\label{appendix C}

In this section, we describe the details of the numerical calculation of the distribution function $p_2(w)$ (\ref{p_n}) for the quantum-state statistics,
and demonstrate additional numerical results for the one-dimensional hard-core boson model (\ref{hardcore boson}).
We also show some results for the one-dimensional spinless fermion model.

The distribution function $p_2(w)$ is numerically calculated by the use of exact diagonalization.
If all the eigenenergies and eigenstates for the initial and final Hamiltonians are known, then
it is straightforward to calculate $p_2(w)$ through the expression (\ref{suppl:p_n}).
In practice, we replace the delta function $\delta(w)$ in Eq.~(\ref{suppl:p_n}) with a rectangular function with a finite grid size $\Delta w$,
\begin{align}
\delta(w)
&=
\begin{cases}
\frac{1}{\Delta w} & 
\tilde w\in [-\frac{\Delta w}{2}, \frac{\Delta w}{2}]
\\
0 & \mbox{otherwise}
\end{cases}.
\end{align}
In the results shown in Fig.~1, we use $\Delta w=0.04$.

To calculate the fluctuation $\Delta p_2$ [Eq.~(\ref{Delta p_2})] for the distribution $p_2(w)$,
we assume the non-degeneracy condition defined by
\begin{align}
&
E_{f,n}-E_{i,m}+E_{f,l}-E_{i,k}
=
E_{f,n'}-E_{i,m'}+E_{f,l'}-E_{i,k'}
\notag
\\
&\Rightarrow\quad
[(k,m)=(k',m') \mbox{ or } (m',k')]
\mbox{ and }
[(l,n)=(l',n') \mbox{ or } (n',l')].
\label{suppl:non-degeneracy}
\end{align}
For the one-dimensional hard-core boson model (\ref{hardcore boson}), there is a trivial degeneracy due to the parity
and translational symmetries. 
There might be other accidental degeneracies in the model.
We assume that these degeneracies can be removed by an infinitesimal perturbation
to the Hamiltonian (\ref{hardcore boson}). 

If the non-degeneracy condition (\ref{suppl:non-degeneracy}) is satisfied, then $\Delta p_2$ can be evaluated as
\begin{align}
\Delta p_2
&=
\frac{1}{Z_i(\beta)}\frac{1}{\mathcal N_2}
\sum_{k,l,m,n}
p_k^i p_m^i
\Big|{\rm Re}[
\langle E_n^f|\hat U|E_m^i\rangle
\langle E_m^i|\hat U^\dagger|E_l^f\rangle \langle E_l^f|\hat U|E_k^i\rangle
\langle E_k^i|\hat U^\dagger|E_n^f\rangle]\Big|,
\end{align}
where the system size $L$ is fixed while $\Delta w\to 0$.
We consider the case in which the Hamiltonian is quenched (i.e., $\hat H=\hat H_i\to\hat H_f$).
In this case,
\begin{align}
\Delta p_2
&=
\frac{1}{Z_i(\beta)}\frac{1}{\mathcal N_2}
\sum_{k,l,m,n}
p_k^i p_m^i
\Big|{\rm Re}[
\langle E_n^f|E_m^i\rangle \langle E_m^i|E_l^f\rangle
\langle E_l^f|E_k^i\rangle \langle E_k^i|E_n^f\rangle]\Big|.
\end{align}
We further assume that the Hamiltonian is real (i.e., $\hat H^\ast=\hat H$), where the eigenstates can be taken as
real vectors. This allows us to rewrite $\Delta p_2$ as
\begin{align}
\Delta p_2
&=
\frac{1}{Z_i(\beta)}\frac{1}{\mathcal N_2}
\sum_{k,l,m,n}
p_k^i p_m^i
|\langle E_n^f|E_m^i\rangle| \cdot |\langle E_m^i|E_l^f\rangle| \cdot
|\langle E_l^f|E_k^i\rangle| \cdot |\langle E_k^i|E_n^f\rangle|.
\end{align}
In the case of $\beta=0$, the expression is further simplified to
\begin{align}
\Delta p_2
&=
\frac{1}{Z_i(0)^2}
\sum_{k,l,m,n}
|\langle E_n^f|E_m^i\rangle| \cdot |\langle E_m^i|E_l^f\rangle| \cdot
|\langle E_l^f|E_k^i\rangle| \cdot |\langle E_k^i|E_n^f\rangle|.
\label{suppl:Delta p_2}
\end{align}
We use Eq.~(\ref{suppl:Delta p_2}) to evaluate $\Delta p_2$ numerically.
For the one-dimensional hard-core boson model, the energy eigenstates in Eq.~(\ref{suppl:Delta p_2}) are taken to be 
simultaneous eigenstates of $\hat P$ and $\hat T+\hat T^{-1}$ \cite{Sandvik2010},
where $\hat P$ is the parity transformation, and $\hat T$ represents the translation to the right by one site.
Note that $\hat H(s)$ (\ref{hardcore boson}), $\hat P$, and $\hat T+\hat T^{-1}$ all commute with each other.

\begin{figure}[t]
\includegraphics[width=10cm]{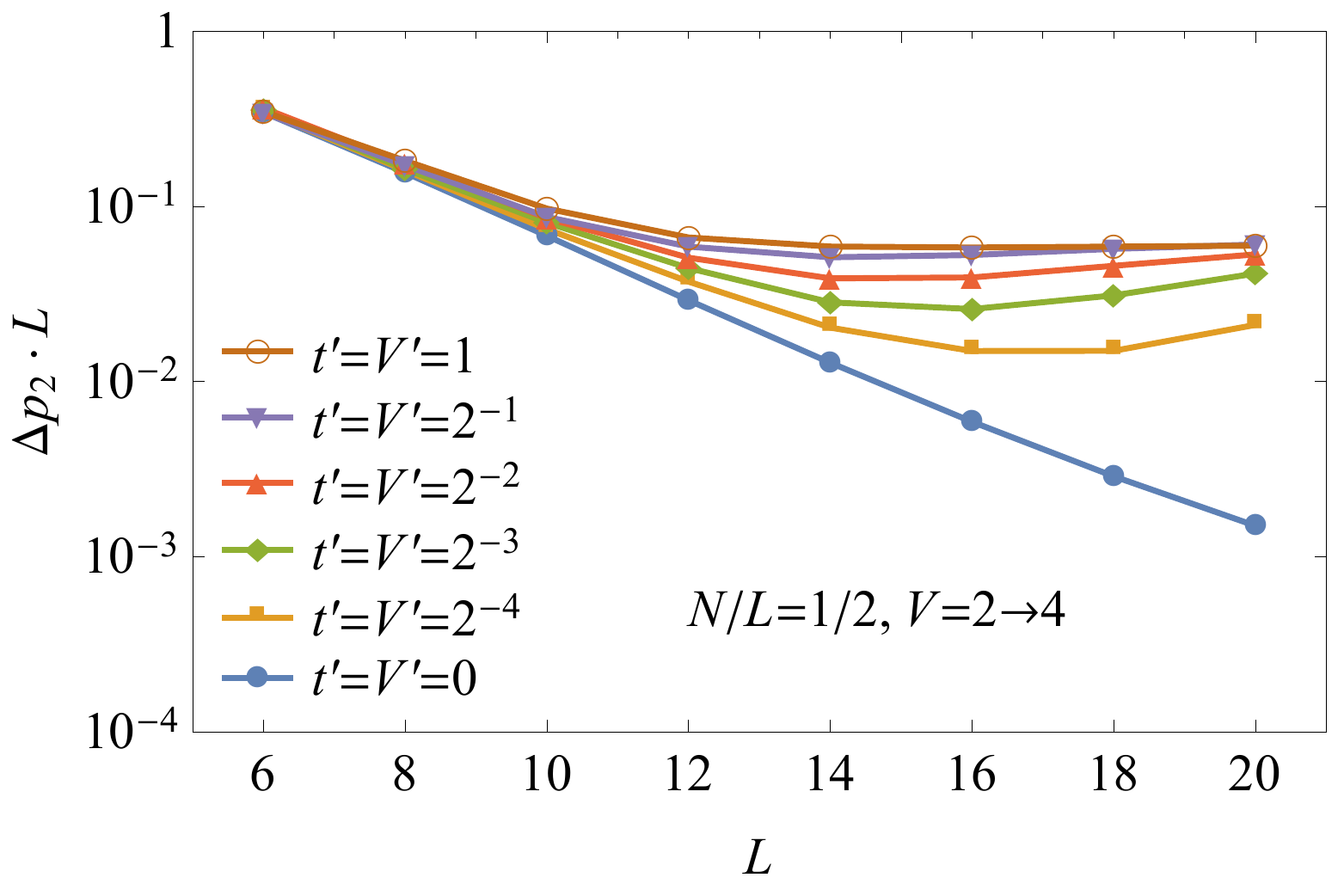}
\caption{The log plot of $\Delta p_2\cdot L$ as a function of $L$ 
for the one-dimensional hard-core boson model driven by the interaction quench $V=2\to 4$
with $\beta=0$ at half filling ($N/L=1/2$).}
\label{suppl:half filling}
\end{figure}

In Fig.~\ref{suppl:half filling}, we plot $\Delta p_2 \cdot L$ as a function of $L$ for the one-dimensional hard-core boson model (\ref{hardcore boson}) at half filling
($N/L=1/2$). By comparing with Fig.~2 (for the filling $N/L=1/3$), one can see that the results do not qualitatively change
while the filling is changed. In both cases, $\Delta p_2$ shows different scaling behaviors between non-integrable
and integrable models. 
For the integrable case ($t'=V'=0$), $\Delta p_2$ decays exponentially, $\Delta p_2\sim e^{-cL}$ with $c=0.36$.
The value of $c$ is different from that for $N/L=1/3$ (shown in the main text), so that $c$ is a non-universal quantity.
On the other hand, in the non-integrable cases ($t'\neq 0$ or $V'\neq 0$) 
the results in Fig.~\ref{suppl:half filling} suggests that $\Delta p_2$ decays algebraically as $\Delta p_2\sim L^{-\gamma}$ with $\gamma=1$.
This supports our expectation that the non-integrable scaling behavior is universal,
and does not depend on details of the system such as the filling. 

\begin{figure}[t]
\includegraphics[width=10cm]{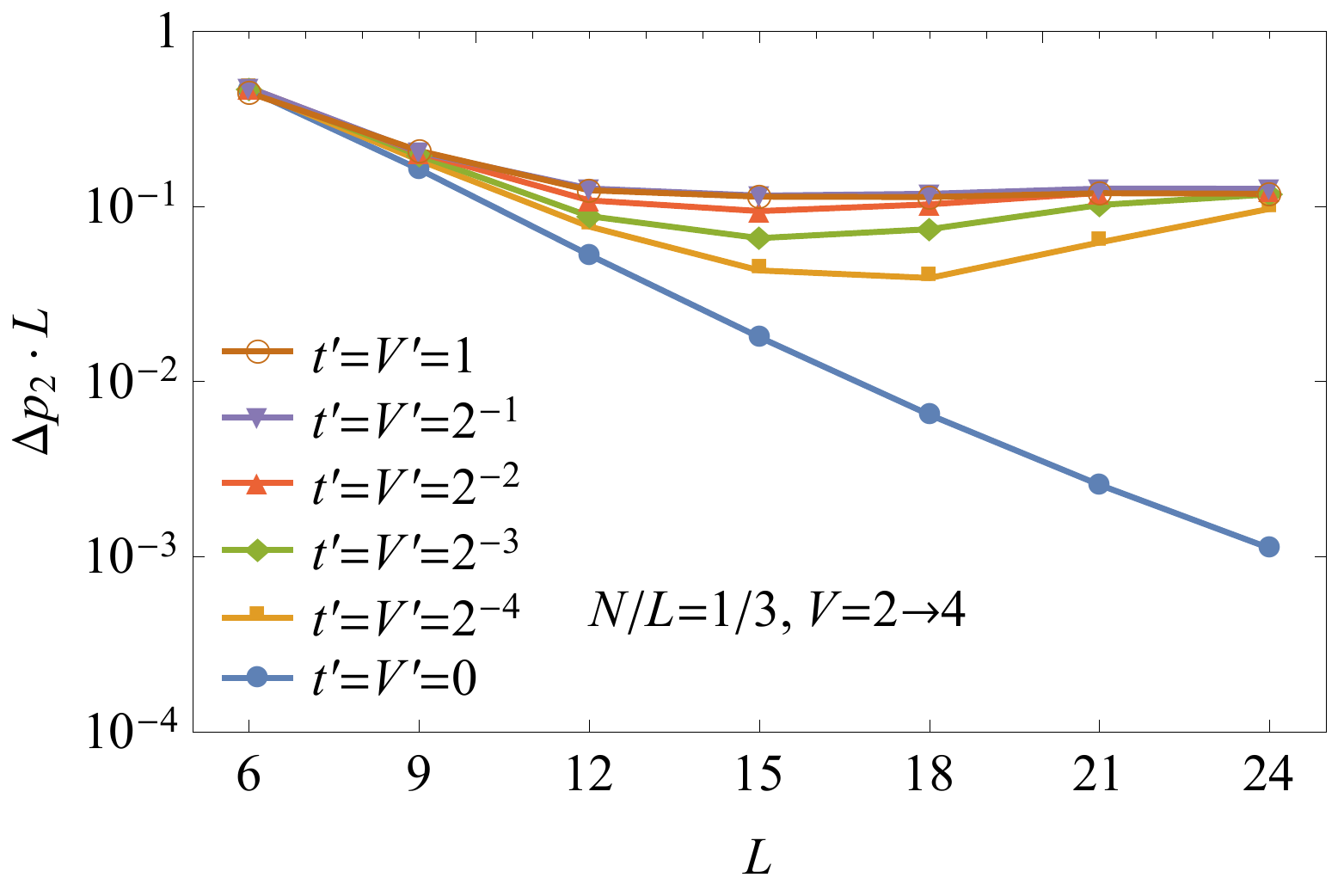}
\caption{The log plot of $\Delta p_2\cdot L$ as a function of $L$
for the one-dimensional spinless fermion model (\ref{suppl:spinless fermion})
driven by the interaction quench $V=2\to 4$ with $\beta=0$ and $N/L=1/3$.}
\label{suppl:fermion weight}
\end{figure}

We also consider the one-dimensional spinless fermion model with nearest and next nearest neighbor hoppings and interactions
\cite{Rigol2009b,SantosRigol2010},
\begin{align}
\hat H(s)
&=
-t\sum_i (f_i^\dagger f_{i+1}+\mbox{h.c.})+V(s)\sum_i n_i^f n_{i+1}^f
-t'\sum_i (f_i^\dagger f_{i+2}+\mbox{h.c.})+V'\sum_i n_i^f n_{i+2}^f,
\label{suppl:spinless fermion}
\end{align}
where $f_i^\dagger$ creates a fermion at site $i$ and $n_i^f\equiv f_i^\dagger f_i$ is the fermion density operator.
The model is known to be integrable when $t'=V'=0$ and non-integrable otherwise \cite{Rigol2009b,SantosRigol2010}.
In Fig.~\ref{suppl:fermion weight}, we plot $\Delta p_2$ as a function of $L$ for the spinless fermion model.
All the non-integrable cases flow into a single universal scaling behavior $\Delta p_2 \sim L^{-\gamma}$ 
with the exponent $\gamma=1$,
which is identical to that for the boson model. On the other hand,
the integrable case shows an exponential decay $\Delta p_2\sim e^{-cL}$ with $c=0.29$.

\end{widetext}

\bibliographystyle{apsrev}
\bibliography{ref}

\end{document}